\documentclass[12pt,onecolumn,journal]{IEEEtran}

\usepackage{amsthm, amssymb}
\newtheorem{thm}{Theorem}

\newtheorem{lem}{Lemma}
\newtheorem{pro}{Proposition}

\newtheorem{rem}{Remark}
\newtheorem{defi}{Definition}

\usepackage{cite}

\usepackage{amsfonts}
\usepackage{multicol,multienum}
\usepackage{subfigure}

\usepackage{color}
\ifCLASSINFOpdf
\else
  \usepackage{float}
  \usepackage[dvips]{graphicx}
  \graphicspath{{../eps/}}
  \DeclareGraphicsExtensions{.eps}
\fi
\usepackage[cmex10]{amsmath}
\usepackage{algorithm}
\usepackage{algorithmic}

\usepackage{array}

\usepackage{setspace}

\usepackage{url}

\begin{document}
%
\title{Learning Equilibrium Play for Stochastic Parallel Gaussian Interference Channels}

\author{Xingqin~Lin
       and~Tat~M.~Lok
\thanks{Xingqin Lin and T. M. Lok are with the Department
of Information Engineering, The Chinese University of Hong Kong, Shatin, N.T., 
Hong Kong. E-mail: \{lxq009, tmlok\}@ie.cuhk.edu.hk.}
}

%


\maketitle

\begin{abstract}
Distributed power control for parallel Gaussian interference channels recently draws great interests. However, all existing works only studied this problem under deterministic communication channels and required certain perfect information to carry out their proposed algorithms. In this paper, we study this problem for stochastic parallel Gaussian interference channels. In particular, we take into account the randomness of the communication environment and the estimation errors of the desired information, and thus formulate a stochastic noncooperative power control game. We then propose a stochastic distributed learning algorithm SDLA-I to help communication pairs learn the Nash equilibrium. A careful convergence analysis on SDLA-I is provided based on stochastic approximation theory and projected dynamic systems approach. We further propose another learning algorithm SDLA-II by including a simple iterate averaging idea into SDLA-I to improve algorithmic convergence performance. Numerical results are also presented to demonstrate the performance of our algorithms and theoretical results.  
\end{abstract}


%
\IEEEpeerreviewmaketitle

\section{Introduction}
The interference channel has long drawn interests from both information theory and communication communities \cite{RefWorks:Cover}. Indeed, the interference channel provides a good model for many communication systems from digital subscriber lines to wireless communication systems. Nevertheless, its capacity region is still unknown in general even in the Gaussian scenario. Moreover, compared to the flat interference channel, fewer works have been done in frequency-selective interference channels. We refer to \cite{RefWorks:vanderMeulen} for an overview on interference channels. 

In this paper we focus on power control in frequency-selective interference channels with Gaussian noise, i.e., parallel Gaussian interference channels. It has been shown recently in \cite{Refworks:ZQLuo22} that obtaining globally optimal solution to maximizing the network sum rates is NP-hard in general. Nevertheless, a distributed game-theoretic approach originally proposed in \cite{RefWorks:WeiYu} becomes increasingly popular. The key assumption is that each individual communication pair is only interested in its own signal and simply treats interference as noise when decoding, i.e, not allowing joint encoding/decoding and interference cancellation techniques. Obviously, this approach provides an inner bound of the capacity region of parallel Gaussian interference channels. More importantly, this strategy is very appealing in practice due to the simplicity and distributiveness. Indeed, receivers in current practical communication systems generally treat interference as noise though substantial research works have been carried out on interference-aware receivers and significant performance gains are promised by multi-user techniques \cite{Refworks:Jeffrey1}.

After the seminal work \cite{RefWorks:WeiYu}, different approaches have been applied to study the distributed power control in parallel Gaussian interference channels when the channel power gains are deterministic. Specifically, \cite{RefWorks:Scutari1} \cite{RefWorks:Scutari2} \cite{RefWorks:Scutari3} are based on contraction mapping, \cite{Refworks:KKL} is based on piecewise affine mapping, \cite{RefWorks:JSPang} resorts to variational inequality theory, and \cite{RefWorks:ZQLuo2} formulates an equivalent linear complementary problem. These works focused on characterizing the Nash equilibrium (NE) such as existence and uniqueness and devising distributed algorithms along with convergence analysis. Indeed, the proposed iterative water-filling algorithm (IWFA) has become a popular candidate for distributed power control in parallel Gaussian interference channels. 

Nevertheless, a common assumption in existing works is that a communication pair is just interested in maximizing its immediate transmission rate. Besides, it is assumed that communication channels remain unchanged during the algorithmic iterations \cite{RefWorks:WeiYu} \cite{RefWorks:Scutari1} \cite{RefWorks:Scutari2}  \cite{RefWorks:Scutari3} \cite{RefWorks:JSPang} \cite{RefWorks:ZQLuo2}. However, the communication time scale is usually large in common applications such as video transmission in wireless data networks \cite{RefWorks:vanderSchaar}. During the whole communication period, it is unlikely that channels would remain the same. In these scenarios, a communication pair may be more interested in maximizing its long term transmission rate rather than the immediate one. Besides, existing works require the knowledge of exact CSI and/or interference levels to be fed back to the corresponding transmitters during the algorithmic iterations. Unfortunately, none of these can be easily obtained in practical communication systems if not impossible. The convergence results of existing schemes such as IWFA are no longer valid or at least unknown when relevant estimation errors exist.

In this paper we take into account the randomness of the communication environment and estimation errors of the desired information. We assume each communication pair is concerned about the long term transmission rate, i.e., the expected transmission rate. We first propose a basic stochastic distributed learning algorithm SDLA-I to help distributed communication pairs learn the NE in stochastic transmission environments. The desired information in implementing SDLA-I is also allowed to be subject to errors. A careful convergence analysis on SDLA-I is also provided based on stochastic approximation theory  \cite{RefWorks:Robbins2} \cite{RefWorks:Harold} and projected dynamic systems (PDS) approach \cite{Refworks:Anna}.  Inspired by the recent developments in stochastic approximation theory \cite{RefWorks:BTPolyak} \cite{RefWorks:Kushner}, we propose another learning algorithm SDLA-II by including a simple iterate averaging idea into the basic learning algorithm SDLA-I to improve the algorithmic convergence performance.

The power control algorithms proposed in this paper belong to the class of stochastic power control algorithms. Existing stochastic power control algorithms (see, e.g., \cite{RefWorks:Ulukus} \cite{RefWorks:Varanasi} \cite{RefWorks:WSWong} and references therein) cannot be applied to the parallel Gaussian interference channels considered in this paper. Note that the recent work \cite{RefWorks:VLau} studied the distributed power control for time-varying parallel Gaussian interference channels. Nevertheless, the model formulated in \cite{RefWorks:VLau} is essentially a deterministic one. So IWFA could still be applicable in \cite{RefWorks:VLau}. In contrast, as explained in section \ref{sec:system}, it would be extremely difficult and/or inconvenient to apply IWFA in our model if not impossible.  Besides, \cite{RefWorks:VLau} also requires the knowledge of exact CSI and interference levels to be fed back to the corresponding transmitters during each iteration of power update. 

The rest of this paper is organized as follows. Section \ref{sec:system} describes the specific system model and the problem formulation. In section \ref{sec:SDLA-I}, the basic learning algorithm SDLA-I is described along with a careful convergence analysis. The PDS approach is adopted in section \ref{sec:PDS} to study the rate of convergence of SDLA-I. We further include the idea of iterate averaging and propose SDLA-II in section \ref{sec:averaging}. Section \ref{sec:numerical} presents some numerical results, and is followed by the conclusions in section \ref{sec:conclusion}.

\section{System Model and Problem Formulation}
\label{sec:system}

\subsection{System Model}

We consider a scenario consisting of a set of $N$ source-destination pairs indexed by $\mathcal{N}=\{1,2,...,N\}$. These communication pairs share a common set $\mathcal{K}=\{1,2,...,K\}$ of frequency-selective unit-bandwidth channels so that their transmissions may interfere with each other. Specifically, the received signal at destination $j$ on the $k$-th channel can be described by the baseband signal model
\begin{align}
y_j^{k} =h_{jj}^{k} \sqrt{p_{j}^{k}} x_{j}^{k} + \sum_{i \neq j, i\in \mathcal{N}} h_{ji}^{k} \sqrt{p_{i}^{k}} x_{i}^{k} + z_j^{k},
\end{align}
where $h_{ji}^{k}$ denotes the channel coefficient from source $i$ to destination $j$ on the $k$-th channel, $p_{j}^k$ denotes the transmission power used by source $j$ on the $k$-th channel, $x_{j}^k$ denotes the normalized transmission symbol of source $j$ on the $k$-th channel, and $z_{j}^k$ denotes the white Gaussian noise with variance $n^k_j$ at destination $j$ on the $k$-th channel.

For later use, we let $g_{ji}^{k} = | h_{ji}^{k} |^2$. In time-varying communication scenarios, channel coefficients are obviously random variables. We denote by $\boldsymbol{G}$ the random vector composed of all the random channel power gain coefficients, i.e., $G_{ji}^{k}, \forall  k \in \mathcal{K}, \forall j, i \in \mathcal{N}$. For the sake of greater applicability we shall make no assumption on the specific underlying statistical distribution of $\boldsymbol{G}$. We simply assume that $\boldsymbol{G}$ is bounded almost surely and different realizations $\boldsymbol{g}$'s of $\boldsymbol{G}$ are independent and identically distributed (i.i.d.). This i.i.d. assumption on $\boldsymbol{G}$ is reasonable in large scale networks.

We further assume that each user is only interested in its own signal and treats interference as noise. Thus, we can write the signal-to-interference-plus-noise-ratio (SINR) at destination $j$ on the $k$-th channel with realization $\boldsymbol{g}$ as
\begin{equation}
\gamma_j^k = \frac{g_{jj}^k p_j^k}{\sum_{i\neq j, i \in \mathbb{N}} g^k_{ji}p_i^k + n_j^k}, \forall  k \in \mathcal{K}, \forall  j \in \mathcal{N}.
\label{eq:3}
\end{equation}
The corresponding maximum achievable rate $R_j$ for user $j$ is given by Shannon formula \cite{RefWorks:Cover}
\begin{align}
R_j (\boldsymbol{p_{j}}, \boldsymbol{p_{-j}} | \boldsymbol{g}) =\sum_{k=1}^{K} \ln (1 + \gamma_j^k),
\label{eq:4}
\end{align}
where $\boldsymbol{p_j}=[p^1_j,p^2_j,...,p^K_j]^T$ denotes the power allocation strategy of user $j$, and $\boldsymbol{p_{-j}}$ denotes the power allocation strategies of all the other users. The power allocation strategy of each user should satisfy certain constraints. Specifically, $\boldsymbol{p_j}$ is regulated by spectral mask constraints, i.e., $0 \leq p^k_j \leq \bar{p}^k_j$, as well as a total power constraint, i.e., $\sum_{k \in \mathcal{K}} p^k_j \leq p_j^{max}$. In order to avoid trivial cases, we assume for all $j \in \mathcal{N}$ that $
\bar{p}^k_j < p_j^{max}, \forall k \in \mathcal{K}, \  \textrm{and} \ \ p_j^{max} < \sum_{k \in \mathcal{K}}\bar{p}^k_j$.

\subsection{Game Theoretical Formulation}

We now formulate the following noncooperative game to characterize the interaction among the users in question:
\begin{equation}
\mathcal{G} = \{\mathcal{N}, \{\Phi_{j}\}_{j \in \mathcal{N}}, \{\bar{R}_j (\boldsymbol{p_{j}}, \boldsymbol{p_{-j}})\}_{j \in \mathcal{N}} \}.
\label{eq:5}
\end{equation}
In game $\mathcal{G}$, $\mathcal{N}$ is the set of players, i.e., communication pairs.  $\bar{R}_j (\boldsymbol{p_{j}}, \boldsymbol{p_{-j}})$ is the utility function of user $j$ given by 
\begin{equation}
\bar{R}_j (\boldsymbol{p_{j}}, \boldsymbol{p_{-j}}) = \mathbb{E}_{\boldsymbol{G}} [R_j (\boldsymbol{p_{j}}, \boldsymbol{p_{-j}} | \boldsymbol{G})],
\label{eq:6}
\end{equation}
where $\mathbb{E}_{\boldsymbol{G}} [\cdot]$ denotes the expected value with respect to $\boldsymbol{G}$. We shall in the sequel drop the subscript to write $\mathbb{E} [\cdot]$ instead of $\mathbb{E}_{\boldsymbol{G}} [\cdot]$ when not leading to confusion. Here we implicitly assume that $\bar{R}_j (\boldsymbol{p_{j}}, \boldsymbol{p_{-j}})$ exists. We further assume $\bar{R}_j (\boldsymbol{p_{j}}, \boldsymbol{p_{-j}})$ is continuous with respect to $\boldsymbol{p}$. $\Phi_{j}$ is the strategy space of user $j$ defined as
\begin{equation}
\Phi_{j} = \{\boldsymbol{p_{j}} \in \mathbb{R}^K: \sum_{k \in \mathcal{K}} p^k_j \leq p_j^{max}, 0 \leq p^k_j \leq \bar{p}^k_j, \forall k \in \mathcal{K} \}.
\label{eq:7}
\end{equation}
For later use, we denote by $\Phi$ the product space $\Phi_{1}\times...\times \Phi_{N}$.

Due to the uncertainty of channel power gains, player $j$ in stochastic game $\mathcal{G}$ wishes to maximize its expected transmission rate $\bar{R}_j$ by choosing appropriate power allocation strategy $\boldsymbol{p_{j}}$. Mathematically, player $j$ solves the following optimization problem 
\begin{align}
\textrm{maximize} \ \ \ \ \  & \bar{R}_j (\boldsymbol{p_{j}}, \boldsymbol{p_{-j}}) \notag \\
\textrm{subject to} \ \ \ \ \ & \boldsymbol{p_{j}} \in \Phi_{j} \notag
\end{align}
where $\bar{R}_j (\boldsymbol{p_{j}}, \boldsymbol{p_{-j}})$ and $\Phi_{j}$ are given in (\ref{eq:6}) and  (\ref{eq:7}), respectively. Note that this is a stochastic optimization problem \cite{RefWorks:Boyd}.

We are interested in understanding if and how the players in stochastic game $\mathcal{G}$ can achieve NE, which is a widely adopted rational outcome of noncooperative games. We formally define NE of the stochastic power control game $\mathcal{G}$ as follows.
\begin{defi} A power allocation profile $\boldsymbol{p^*} = (\boldsymbol{p_1^*},...,\boldsymbol{p_N^*} )$ is called an NE of the stochastic power control game $\mathcal{G}$ if and only if 
\begin{align}
\boldsymbol{p_j^*} \in \textrm{arg}\max \{\bar{R}_j (\boldsymbol{p_{j}}, \boldsymbol{p^*_{-j}}): \boldsymbol{p_{j}} \in \Phi_{j}  \}, \forall j \in \mathcal{N}.
\label{eq:8}
\end{align}
\end{defi} 
Game $\mathcal{G}$ has been extensively studied when the channel power gains are deterministic. Nevertheless, new challenges arise due to the randomness in the channel power gains caused by the stochastic communication environments. Indeed, player $j$ in stochastic game $\mathcal{G}$ may not even be able to know its utility function $\mathbb{E} [R_j (\boldsymbol{p_{j}}, \boldsymbol{p_{-j}} | \boldsymbol{G})]$ due to the following reasons. Firstly, the distribution of $\boldsymbol{G}$ is unknown though $R_j (\boldsymbol{p_{j}}, \boldsymbol{p_{-j}}| \boldsymbol{g})$ is known. Thus, it is impossible to evaluate $\mathbb{E} [R_j (\boldsymbol{p_{j}}, \boldsymbol{p_{-j}}| \boldsymbol{G})]$ analytically or numerically. Indeed, even if the distribution of $\boldsymbol{G}$ was known, it would require player $j$ to obtain global knowledge to evaluate $\mathbb{E} [R_j (\boldsymbol{p_{j}}, \boldsymbol{p_{-j}}| \boldsymbol{G})]$, which may result in an unacceptable level of communication overhead. Furthermore, even further assuming that player $j$ has the global knowledge about the distribution of $\boldsymbol{G}$, evaluation of $\mathbb{E} [R_j (\boldsymbol{p_{j}}, \boldsymbol{p_{-j}}| \boldsymbol{G})]$ involves multi-dimensional integration and is thus computationally expensive. Since player $j$ does not know its exact utility function $\mathbb{E} [R_j (\boldsymbol{p_{j}}, \boldsymbol{p_{-j}}| \boldsymbol{G})]$, it is impossible for player $j$ to compute a best response, which is an essential component in IWFA. So IWFA cannot be applied to the stochastic game $\mathcal{G}$ investigated in this paper.

\section{Stochastic Algorithm for Learning NE}
\label{sec:SDLA-I}

\subsection{Stochastic Distributed Learning Algorithm I}

We aim to design a distributed scheme so that an NE of the stochastic game $\mathcal{G}$ can be obtained even with so many difficulties described in the previous section. Obviously, such a distributed scheme makes sense only when NE exists. Thus, we first address the existence of NE in the following proposition.

\begin{pro} 
At least one NE exists in the stochastic power control game $\mathcal{G}$.
\end{pro}
\begin{proof}
See Appendix \ref{proof:pro1}.
\end{proof}

In stochastic communication environments, a desired distributed scheme must offer users time to ``learn'' the environments gradually. Hopefully, an NE can be achieved as users in game $\mathcal{G}$ keep taking adaptive strategies during the learning process. Toward this end, we first define $f^k_j$ as
\begin{align}
f^k_j = \frac{g^k_{jj}p_j^k}{\sum_{i \in \mathcal{N}} g^k_{ji}p_i^k + n_j^k},
\end{align} 
which represents the ratio of the received energy of user $j$'s signal to the total received signal energy at destination $j$ on the $k$-th channel. We let $\boldsymbol{f_j} = [f^1_j,...,f^K_j]^T$. Since $R_j (\boldsymbol{p_{j}}, \boldsymbol{p_{-j}} | \boldsymbol{g})$ is concave with respect to $\boldsymbol{p_{j}}$ and $\boldsymbol{f_j}/\boldsymbol{p_{j}}$ is the associated gradient, we have for any $\boldsymbol{p_{j}} \in \Phi_j$ 
\begin{align}
R_j (\boldsymbol{q_{j}}, \boldsymbol{p_{-j}} | \boldsymbol{g}) \leq R_j (\boldsymbol{p_{j}}, \boldsymbol{p_{-j}} | \boldsymbol{g}) + (\boldsymbol{f_j}/\boldsymbol{p_{j}})^T (\boldsymbol{q_{j}} - \boldsymbol{p_{j}}), \forall \boldsymbol{q_{j}} \in \Phi_j. \notag
\end{align}

Now we are in a position to describe the distributed learning algorithm SDLA-I for game $\mathcal{G}$ to reach an NE. We formally summarize SDLA-I in Table \ref{alg2}. 

\begin{algorithm}
\caption{SDLA-I}
\label{alg2}
\algsetup{indent=2em}
\begin{algorithmic}
\STATE \textbf{Step 1: Initialization:}\\
Each player $j \in \mathcal{N}$ starts with an arbitrarily feasible power allocation vector, i.e., $\boldsymbol{p_j} (0) \in \Phi_j$. Set $n:=0$. 
\STATE \textbf{Step 2: Computation:}\\
Each player $j\in \mathcal{N}$ computes $\boldsymbol{p_j} (n+1)$ by
\begin{align}
\boldsymbol{p_j} (n+1) = \mathcal{P}_{\Phi_j} [\boldsymbol{p_j} (n) + a_j(n) \frac{\boldsymbol{\hat{f}_j} (n)}{\boldsymbol{p_j} (n)}],
\label{eq:9}
\end{align}
where $\mathcal{P}_{\Phi_j} [\cdot]$ denotes the projection onto $\Phi_j$ with respect to the Euclidean norm, $(a_j(n))_{n=0}^{\infty}$ is step size sequence.
\STATE \textbf{Step 3: Convergence Verification:}\\
If stopping criteria are satisfied, then stop; otherwise, set $n:=n+1$, and go to Step 2.
\end{algorithmic}
\end{algorithm}

In equation (\ref{eq:9}), $\boldsymbol{\hat{f}_j} (n) = (\hat{f}^1_j(n),...,\hat{f}^K_j(n))^T$ where $\hat{f}^k_j(n)$ is an approximate estimate of $f^k_j(n+1)$. Thus, receiver $j$ can just locally measure the total received signal energy and extract its own signal energy on each subchannel. Then receiver $j$ notifies transmitter $j$ through control channel the corresponding ratio vector $\boldsymbol{\hat{f}_j}$. Note that SDLA-I does not require an exact estimate. Mathematically, 
\begin{align}
R_j (\boldsymbol{p_{j}}, \boldsymbol{p_{-j}}(n) | \boldsymbol{g} (n+1)) \leq & R_j (\boldsymbol{p_{j}}(n), \boldsymbol{p_{-j}}(n) | \boldsymbol{g}(n+1))   \notag \\
&+(\boldsymbol{\hat{f}_j}(n)/\boldsymbol{p_{j}}(n))^T (\boldsymbol{p_{j}} - \boldsymbol{p_{j}}(n)) + \epsilon_j (n), \forall \boldsymbol{p_{j}} \in \Phi_j, 
\label{eq:10}
\end{align}
where $\epsilon_j \geq 0$ measures the accuracy of the estimation $\boldsymbol{\hat{f}_j}$. Note that all existing algorithms for distributed power control in parallel Gaussian interference channels require the knowledge of exact CSI and/or interference level to be fed back to the corresponding transmitters\cite{RefWorks:WeiYu} \cite{RefWorks:Scutari1} \cite{RefWorks:Scutari2}  \cite{RefWorks:Scutari3} \cite{RefWorks:JSPang} \cite{RefWorks:ZQLuo2} \cite{RefWorks:VLau}. Unfortunately, it is hard to obtain perfect knowledge of these information in practical communication systems if not impossible. The convergence results on existing schemes such as IWFA are no longer valid or at least unknown when relevant estimation errors exist.  Thus, as described above, our proposed SDLA-I is more robust and requires less communication overhead. Nevertheless, the estimation errors of $\boldsymbol{\hat{f}_j}$ should not be too ``bad''. We later will formalize the quantitative criteria which specify how exact $\boldsymbol{\hat{f}_j}$ should be.

A careful reader may concern about the computation complexity of SDLA-I since each player needs to conduct a projection operation during every iteration and projection is in general time-consuming. We address this issue through the following proposition which in fact provides a closed form solution for the projection operation in (\ref{eq:9}), implying that SDLA-I can be carried out efficiently.

\begin{pro} 
The closed form solution for the projection operation (\ref{eq:9}) is given by 
\begin{equation}
p_{j}^{k} (n+1)=[p_j^k (n) + a_j(n)\frac{\hat{f}_{j}^k(n)}{p_{j}^k(n)}  - \lambda_j]_0^{ \bar{p}_{j}^{k}}, \forall k \in \mathcal{K},
\label{eq:12}
\end{equation}                                                        
where $[x]_a^b = \max ( a,\min ( x,b )) $, and $\lambda_{j} \geq 0$ is chosen to satisfy $\sum_{k \in \mathcal{K}}p_{j}^{k} (n+1)=p_{j}^{max}$. 
\end{pro}

\begin{proof}
See Appendix \ref{proof:pro2}.
\end{proof}

\subsection{Convergence of SDLA-I}

In this subsection, we study the convergence property of SDLA-I. Toward this end, we first introduce some further notations for ease of exposition. We denote by $D(n) = diag(D_1(n),...,D_N(n))$ the $NK \times NK$-dimensional block diagonal matrix where $D_j(n) = diag(a_j(n),...,a_j(n))$ is a $K \times K$-dimensional diagonal matrix with uniform diagonal entry $a_j(n)$. Then the iteration step (\ref{eq:9}) in SDLA-I can be rewritten in a compact form given by
\begin{align}
\boldsymbol{p} (n+1) = \mathcal{P}_{\Phi} [\boldsymbol{q}(n)],
\label{eq:17}
\end{align}
where $\boldsymbol{q}(n) = \boldsymbol{p}(n) + D(n) \frac{\boldsymbol{\hat{f}}(n)}{\boldsymbol{p}(n)}$ with $\boldsymbol{\hat{f}}(n)=(\boldsymbol{\hat{f}_1}(n),..,\boldsymbol{\hat{f}_N}(n))^T$. Denote by $\boldsymbol{s}_j (n) = \boldsymbol{f_j}(n)/\boldsymbol{p_{j}}(n)$, $\boldsymbol{\hat{s}}_j (n) = \boldsymbol{\hat{f}_j}(n)/\boldsymbol{p_{j}}(n)$ and $\boldsymbol{\bar{s}}_j  = \mathbb{E} [ \nabla_{\boldsymbol{p_j}}  R_j(\boldsymbol{p_{j}}, \boldsymbol{p_{-j}} | \boldsymbol{G}) ]$. We group all the $\boldsymbol{s}_j (n)$'s, $\boldsymbol{\hat{s}}_j (n)$'s, and $\boldsymbol{\bar{s}}_j$'s into column vectors $\boldsymbol{s} (n)$, $\boldsymbol{\hat{s}}(n)$, and $\boldsymbol{\bar{s}}$, respectively. We further denote by $\boldsymbol{\Gamma}$ the $N \times N$-dimensional matrix with $[\boldsymbol{\Gamma}]_{ij}$ defined as
\[
[\boldsymbol{\Gamma}]_{ij} = \left\{\begin{array}{ll}
                 1     &\mbox{if $i=j$,} \\  
                 -\max_{k \in \mathcal{K}} (\frac{\displaystyle g_{ij}^k}{\displaystyle g_{jj}^k} \cdot \frac{\displaystyle n_j^k + \sum_{j^\dagger \in \mathcal{N}} g_{j j^\dagger}^k \bar{p}^k_{j^\dagger}}{\displaystyle n_i^k} )     &\mbox{if $i\neq j$.}
                \end{array} \right.
\]

With these notations in mind, the following lemma summarizes some main (in)equalities, which will be used in the later proofs of the convergence results of SDLA-I.

\begin{lem} 
The following (in)equalities hold:
\begin{itemize}
\item[(i)] A power allocation profile $\boldsymbol{p^*} \in \Phi$ is an NE of the stochastic game $\mathcal{G}$ if and only if for any $a_j >0$
\begin{align}
\boldsymbol{p_j^*} = \mathcal{P}_{\Phi_j} [\boldsymbol{p_j^*} + a_j \frac{\boldsymbol{\bar{f}_j}}{\boldsymbol{p_j^*}} ], \forall j \in \mathcal{N},
\label{eq:16}
\end{align}
where $\boldsymbol{\bar{f}_j} = \boldsymbol{p_j^*} \boldsymbol{\bar{s}}_j (\boldsymbol{p^*})$.
\item[(ii)] For any $\boldsymbol{p}, \boldsymbol{\bar{p}} \in \mathbb{R}^{NK}$,
\begin{align}
\parallel \mathcal{P}_{\Phi} (\boldsymbol{p}) - \mathcal{P}_{\Phi} (\boldsymbol{\bar{p}}) \parallel \leq \parallel \boldsymbol{p} -  \boldsymbol{\bar{p}} \parallel.
\label{eq:1616}
\end{align}
\item[(iii)] For any $\boldsymbol{p} \in \mathbb{R}^{NK}$ and $\boldsymbol{\bar{p}} \in \Phi$,
\begin{align}
(\boldsymbol{\bar{p}} - \boldsymbol{p})^T(\mathcal{P}_{\Phi} (\boldsymbol{p}) - \boldsymbol{p} ) \geq 0.
\label{eq:161616}
\end{align}
\end{itemize}
\end{lem}
\begin{proof}
See Appendix \ref{proof:lem1}.
\end{proof}

The following lemma inspired by \cite{RefWorks:JSPang} provides another inequality (\ref{eq:lem1}) that will be used later. As a byproduct, we also characterize the uniqueness property of NE in game $\mathcal{G}$ with deterministic channel power gains in the following lemma. We refer to \cite{RefWorks:Scutari1} \cite{RefWorks:JSPang} and references therein for a more detail discussion on the uniqueness property of NE in deterministic game $\mathcal{G}$.

\begin{lem}
For given channel power gain realization $\boldsymbol{g}$, if $\boldsymbol{\Gamma} \succ \boldsymbol{0}$ (positive definite), then there exists a unique NE $\boldsymbol{p^*}(\boldsymbol{g}) \in \Phi$, and
\begin{align}
\boldsymbol{s} (\boldsymbol{p} | \boldsymbol{g} )^T (\boldsymbol{p^*}(\boldsymbol{g}) - \boldsymbol{p}) \geq \tau(\boldsymbol{s}) \parallel \boldsymbol{p} - \boldsymbol{p^*}(\boldsymbol{g}) \parallel_2^2, \forall \boldsymbol{p} \in \Phi,
\label{eq:lem1}
\end{align}
with
\begin{align}
\tau(\boldsymbol{s}) = \frac{\displaystyle \lambda_{\min} (\boldsymbol{\Gamma}) }{\displaystyle \max_{i\in \mathcal{N}} \max_{k\in \mathcal{K}} (\kappa^k_i)^2 }
\label{eq:lem11}
\end{align}
where $\lambda_{\min} (\boldsymbol{\Gamma})>0$ denotes the minimal eigenvalue of the symmetric part of $\boldsymbol{\Gamma}$, and $\kappa^k_i =(n_i^k + \sum_{j\in \mathcal{N}} g_{ij}^k \bar{p}_j^k)/g_{ii}^k$.
\end{lem}

\begin{proof}
See Appendix \ref{proof:lem2}.
\end{proof}

We now describe the convergence results of SDLA-I in the following theorem.

\begin{thm}
Let $\mathcal{F}_n$ be the $\sigma$-field generated by $(\boldsymbol{g}(m),\boldsymbol{p}(m))_{m=0}^{n}$. Assume that:
\begin{itemize}
\item[(i)] $\boldsymbol{\Gamma} \succ \boldsymbol{0}$ holds almost surely.
\item[(ii)] The step sizes $a_i (n) \geq 0$ satisfy:
\begin{align}
&\sum_{n=0}^{\infty} \min_{i\in \mathcal{N}} a_i (n) = +\infty, \\
&\sum_{n=0}^{\infty} a^2_i (n) < +\infty, \forall i \in \mathcal{N}.
\end{align}
\item[(iii)] The estimation errors $\epsilon_i (n) \geq 0$ satisfy:
\begin{align}
\sum_{n=0}^{\infty} \mathbb{E}[a_i(n) \epsilon_i (n) | \mathcal{F}_n] < +\infty, \forall i \in \mathcal{N}.
\end{align}
\end{itemize}
Then $(\boldsymbol{p}(n))_{n=0}^{\infty}$ generated by SDLA-I converges to the unique  NE $\boldsymbol{p^*}$ of the stochastic game $\mathcal{G}$ in the mean square sense, i.e., $\lim_{n\rightarrow \infty} \parallel \boldsymbol{p}(n) - \boldsymbol{p^*} \parallel^2 =0$ almost surely.
\end{thm}

\begin{proof}
See Appendix \ref{proof:thm1}.
\end{proof}

We make the following remarks on the assumptions in Theorem 1:

\begin{rem}
Assumption (i) is the major requirement for the convergence of SDLA-I. A careful thinking reveals that this assumption is indeed intuitive. On the one hand, from the game theory point of view, $\boldsymbol{\Gamma} \succ \boldsymbol{0}$ implies that each player $j$ has a more significant influence on its utility than other players do. From the communication point of view, $\boldsymbol{\Gamma} \succ \boldsymbol{0}$ imposes upper bounds on the interference received and/or caused by communication pair $j$. Under mild interference conditions, communication pair $j$'s achievable transmission rate is not heavily influenced by other communication pairs. Note that all existing algorithms even for deterministic distributed power control such as IWFA require more or less similar conditions to ensure convergence \cite{RefWorks:WeiYu} \cite{RefWorks:Scutari1} \cite{RefWorks:Scutari2}  \cite{RefWorks:Scutari3} \cite{RefWorks:JSPang} \cite{RefWorks:ZQLuo2} \cite{RefWorks:VLau}.
\end{rem}

\begin{rem}
Assumption (ii) is quite standard in stochastic approximation algorithms. Indeed, condition (18), i.e., $\sum_{n=0}^{\infty} \min_{i\in \mathcal{N}} a_i (n) = +\infty$, ensures that SDLA-I can cover the entire time axis to reach the NE in stochastic parallel Gaussian interference channels. Meanwhile, the choice of step sizes such that $\sum_{n=0}^{\infty} \mathbb{E} [ a^2_i (n)|\mathcal{F}_n ] < +\infty$ can asymptotically suppress error variance during the learning process. 
\end{rem}

\begin{rem}
Assumption (iii) provides a quantitative answer to the question on how well the estimation $\boldsymbol{\hat{f}}$ should be. Specifically, $\sum_{n=0}^{\infty} \mathbb{E}[a_j(n) \epsilon_j (n) | \mathcal{F}_n] < +\infty$ implies that the total estimation errors can be controlled. This assumption is reasonable especially in slow to medium time-varying communication environments. Nevertheless, the estimation $\boldsymbol{\hat{f}}$ may not be good enough in fast time-varying scenarios. In this regard, better estimates may be required. Noting that $\boldsymbol{\hat{f}}(n)$ only utilizes the last feedback $\boldsymbol{f}(n)$, one possible solution is to take advantage of empirical distribution after having observed many realization $\boldsymbol{f}$'s. That is, distributed communication pairs gradually learn more about the environment. Thus, better estimates $\boldsymbol{\hat{f}}(n)$ may be obtained. Nevertheless, we do not aim to explore this topic which is beyond the scope of this paper. 
\end{rem}

To further appreciate how SDLA-I works, let us consider a particular scenario where the difference between $\boldsymbol{\hat{f}_j}(n)/\boldsymbol{p_j}(n)$ and $\mathbb{E}[\nabla_{\boldsymbol{p_j}} R_j (\boldsymbol{p_j}(n), \boldsymbol{p_{-j}}(n) | \boldsymbol{G})]$ is captured by random vector $\boldsymbol{\theta_j}(n)$, i.e.,
\begin{align}
\boldsymbol{\hat{f}_j}(n)/\boldsymbol{p_j}(n) = \mathbb{E}[\nabla_{\boldsymbol{p_j}} R_j (\boldsymbol{p_j}(n), \boldsymbol{p_{-j}}(n) | \boldsymbol{G})] + \boldsymbol{\theta_j}(n), \forall j \in \mathcal{N}.
\label{eq:244}
\end{align}
As usual, we group all $\boldsymbol{\theta_j}$'s into a column vector $\boldsymbol{\theta}$, i.e., $\boldsymbol{\theta} =(\boldsymbol{\theta_1},...,\boldsymbol{\theta_N})^T$. In other words, we simply use an online estimate  $\boldsymbol{\hat{f}_j}(n)/\boldsymbol{p_j}(n)$ to approximate $\mathbb{E}[\nabla_{\boldsymbol{p_j}} R_j (\boldsymbol{p_j}(n), \boldsymbol{p_{-j}}(n) | \boldsymbol{G})]$ though we are not able to evaluate $\mathbb{E}[\nabla_{\boldsymbol{p_j}} R_j (\boldsymbol{p_j}(n), \boldsymbol{p_{-j}}(n) | \boldsymbol{G})]$. The approximation difference is captured by $\boldsymbol{\theta_j}$. We will show that SDLA-I converges as long as this simple approximation is not too ``bad''. We will formalize these ideas in Theorem 2. Toward this end, we first prove a simple lemma as follows.

\begin{lem}
The mapping $\boldsymbol{\bar{s}}(\boldsymbol{p})$ where $\boldsymbol{\bar{s}_j} (\boldsymbol{p}) = \mathbb{E} [ \nabla_{\boldsymbol{p_j}}  R_j(\boldsymbol{p_{j}}, \boldsymbol{p_{-j}} | \boldsymbol{G}) ] $ is Lipschitz continuous almost surely. That is, there exists a positive constant $L$ such that $\forall \boldsymbol{p}, \boldsymbol{q} \in \Phi$,
\begin{align}
\parallel \boldsymbol{\bar{s}} (\boldsymbol{p}) - \boldsymbol{\bar{s}} (\boldsymbol{q}) \parallel \leq L \parallel \boldsymbol{p} - \boldsymbol{q} \parallel
\label{eq:2444}
\end{align}
holds almost surely.
\end{lem}

\begin{proof}
See Appendix \ref{proof:lem3}.
\end{proof}

\begin{thm}
Let $\mathcal{F}_n$ be the $\sigma$-field generated by $(\boldsymbol{g}(m),\boldsymbol{p}(m))_{m=0}^{n}$. Assume that:
\begin{itemize}
\item[(i)] $\boldsymbol{\Gamma} \succ \boldsymbol{0}$ holds almost surely.
\item[(ii)] The step sizes $a_i (n) \geq 0$ satisfy:
\begin{align}
2 \tau(\boldsymbol{\bar{s}}) \min_{i\in \mathcal{N}} a_{i} (n)  \geq L^2 \max_{i \in \mathcal{N}} a^2_i(n) + \delta(n),
\label{eq:24444}
\end{align}
where $\delta (n)$ is any bounded positive constant. 
\item[(iii)] The difference random vector $\boldsymbol{\theta} (n)$ satisfy:
\begin{align}
&\mathbb{E} [\boldsymbol{\theta}(n) | \mathcal{F}_n] = \boldsymbol{0}, \\
&\sum_{n=0}^{\infty} \sum_{i \in \mathcal{N}} a^2_i (n) \mathbb{E}[ \parallel \boldsymbol{\theta_i}(n)  \parallel^2 | \mathcal{F}_n] < +\infty.
\end{align}
\end{itemize}
Then $(\boldsymbol{p}(n))_{n=0}^{\infty}$ generated by SDLA-I converges to the unique  NE $\boldsymbol{p^*}$ of the stochastic game $\mathcal{G}$ in the mean square sense, i.e., $\lim_{n\rightarrow \infty} \parallel \boldsymbol{p}(n) - \boldsymbol{p^*} \parallel^2 = 0$ almost surely.
\end{thm}

\begin{proof}
See Appendix \ref{proof:thm2}.
\end{proof}

Note that distributed algorithms based on the gradient projection mapping for deterministic parallel Gaussian interference channels have been proposed in \cite{RefWorks:Scutari2}. Nevertheless, the convergence behaviors of those algorithms in \cite{RefWorks:Scutari2} are only shown for deterministic scenarios and thus cannot be applied to stochastic cases. Indeed, Theorem 2 establishes a theoretical foundation for the convergence of those deterministic algorithms under stochastic scenarios. The key conditions are included in assumption (iii) in Theorem 2. That is, the naive estimate  $\boldsymbol{\hat{f}_j}(n)/\boldsymbol{p_j}(n)$ for $\mathbb{E}[\nabla_{\boldsymbol{p_j}} R_j (\boldsymbol{p_j}(n), \boldsymbol{p_{-j}}(n) | \boldsymbol{G})]$ should not be too ``bad'' in the sense of assumption (iii) in Theorem 2.

Besides, the requirement ($\ref{eq:24444}$) imposed on step sizes is also reasonable. Consider a common step size choice for every communication pair, i.e., $a_{i} (n) = \bar{a}(n), \forall i \in \mathcal{N}$. Ignoring the arbitrarily small constant $\delta (n)$ for ease of exposition, condition $(\ref{eq:24444})$ is then reduced to $\bar{a}(n) \leq \frac{2 \tau(\boldsymbol{\bar{s}})}{L^2}, \forall n$. That is, larger step size can be taken if $\boldsymbol{\bar{s}}$ is more strongly monotone (i.e., larger $\tau(\boldsymbol{\bar{s}})$). In contrast, smaller step size should be taken if $\boldsymbol{\bar{s}}$ changes more significantly with respect to $\boldsymbol{p}$ (i.e., larger Lipschitz constant $L$).

Though Theorem 2 is of interest in theory, we remark that assumptions in Theorem 2 may not be easily verified. For instance, it is hard to know if the difference random vector $\boldsymbol{\theta}$ could satisfy assumption (iii) if little is known about the distribution of $\boldsymbol{G}$ in real communication systems. Besides, requirement ($\ref{eq:24444}$) imposed on step sizes involves strongly monotone modulus $\tau(\boldsymbol{\bar{s}})$ and Lipschitz constant $L$, both of which depend on the specific channel gain distribution $\boldsymbol{G}$. In contrast, the step sizes choice in Theorem 1 is relatively standard. The requirement there is that the total error in the stochastic gradient obtained by local communication pair could be properly controlled. This requirement may be easily satisfied when the parallel Gaussian interference channels do not change too fast. 

\section{Continuous Time Approximation by PDS}
\label{sec:PDS}

Note that previous convergence results do not provide insights on the speed of convergence of SDLA-I. Indeed, they may be considered as study of the accuracy of SDLA-I. Equally important is the convergence rate of SDLA-I. In this section, we shall shed some lights on this question. We note that an exact analysis on the convergence rate of SDLA-I is extremely difficult if not impossible due to the various stochastic factors. Therefore, we resort to a PDS approach which approximates but still captures the essential behaviors of SDLA-I to help us appreciate the convergence speed. Note that a PDS formulation for transient behavior analysis for deterministic cognitive radio networks was also briefly described in \cite{RefWorks:Setoodeh}.

To begin with, we recall some basic concepts of PDS from \cite{Refworks:Anna} to facilitate  further discussions. Consider a closed convex set $\boldsymbol{\mathcal{K}} \in \mathcal{R}^M$ and a vector field $\boldsymbol{\mathcal{F}}$ whose domain contains $\boldsymbol{\mathcal{K}}$. Recall $\mathcal{P}_{\boldsymbol{\mathcal{K}}}$ denotes the norm projection. Then define the projection of $\boldsymbol{\mathcal{F}}$ at $\boldsymbol{x}$ as
\begin{equation}
\prod_{\boldsymbol{\mathcal{K}}} (\boldsymbol{x}, \boldsymbol{\mathcal{F}}) = \lim_{\delta\rightarrow 0} \frac{\mathcal{P}_{\boldsymbol{\mathcal{K}}} (\boldsymbol{x} + \delta \boldsymbol{\mathcal{F}} ) - \boldsymbol{x}}{\delta}.
\label{eq:PDS2}
\end{equation}
Now we formally define PDS as follows.
\begin{defi}
The following ordinary differential equations
\begin{equation}
\dot{\boldsymbol{x}} (t) = \prod_{\boldsymbol{\mathcal{K}}} (\boldsymbol{x} (t), -\boldsymbol{\mathcal{F}}(\boldsymbol{x} (t)))
\label{eq:PDS3}
\end{equation}
with an initial value $\boldsymbol{x}(0) \in \boldsymbol{\mathcal{K}}$ is called projected dynamical system $PDS(\boldsymbol{\mathcal{F}},\boldsymbol{\mathcal{K}})$.
\end{defi}

Note that the right hand side in (\ref{eq:PDS3}) is discontinuous on the boundary of $\boldsymbol{\mathcal{K}}$ due to the projection operator, which is different from classical dynamical systems. 

Now let us consider $PDS(\boldsymbol{\bar{s}},\Phi)$ given by $\dot{\boldsymbol{p}} (t) = \prod_{\Phi} (\boldsymbol{p} (t), \boldsymbol{\bar{s}}(\boldsymbol{p} (t)))$ with initial value $\boldsymbol{p}(0) \in \Phi$. The key results of this PDS are summarized in the following proposition.

\begin{pro} 
The $PDS(\boldsymbol{\bar{s}},\Phi)$ with initial value $\boldsymbol{p}(0)$ has the following properties:
\begin{itemize}
\item[(i)] It has a unique solution $\boldsymbol{p} (t)$ which continuously depends on the vector field $\boldsymbol{\bar{s}}$ and initial value $\boldsymbol{p} (0)$;
\item[(ii)] A vector $\boldsymbol{p^*} \in \Phi$ is the NE of the stochastic game $\mathcal{G}$ if and only if it is a stationary point of $PDS(\boldsymbol{\bar{s}},\Phi)$, i.e., $\dot{\boldsymbol{p^*}} (t) = 0$;
\item[(iii)] If $\boldsymbol{\Gamma} \succ \boldsymbol{0}$ holds almost surely, then stationary point $\boldsymbol{p^*}$ of $PDS(\boldsymbol{\bar{s}},\Phi)$ is unique and globally exponentially stable, i.e., $\parallel \boldsymbol{p} (t) - \boldsymbol{p^*} \parallel \leq \parallel \boldsymbol{p}(0) - \boldsymbol{p^*} \parallel  \exp (- \min_{\boldsymbol{s}} \tau (\boldsymbol{s}) \cdot t)$ with $\tau (\boldsymbol{s})$ given in (\ref{eq:lem11}).
\end{itemize}
\end{pro}

\begin{proof}
See Appendix \ref{proof:pro3}.
\end{proof}

$PDS(\boldsymbol{\bar{s}},\Phi)$ is the underlying idealized version of SDLA-I. In other words, we can view SDLA-I as a stochastic approximation of $PDS(\boldsymbol{\bar{s}},\Phi)$ \cite{Refworks:Anna}. Thus, the iteration process $(\boldsymbol{p}(n))_{n=0}^{\infty}$ in SDLA-I approximates or tracks the solution $\boldsymbol{p}(t)$ of $PDS(\boldsymbol{\bar{s}},\Phi)$. From the above proposition, we know that $\boldsymbol{p}(t)$ converges to $\boldsymbol{p^*}$ at an exponential rate. Note that the stationary point $\boldsymbol{p^*}$ of $PDS(\boldsymbol{\bar{s}},\Phi)$ is also the limit point of $(\boldsymbol{p}(n))_{n=0}^{\infty}$. So we can expect that $\boldsymbol{p}(n)$ moves in an approximately (subject to the inherent stochastic variations) monotone fashion to $\boldsymbol{p^*}$ at an exponential rate. This understanding of the iteration process in SDLA-I is also instrumental in exploring the idea of iterate averaging, which is detailed in the next section.

\section{Learning NE with Averaging}
\label{sec:averaging}

Fast convergence performance of distributed learning algorithm for obtaining an NE of the power control game $\mathcal{G}$ is clearly desirable in real communication systems. From previous discussions, we can see that the choice of good step sizes $a_i (n)$ has a profound effect on the convergence performance of SDLA-I. In this section, we discuss how we can improve the convergence performance of SDLA-I so that distributed communication pairs can learn the NE in a faster fashion. 

Among various approaches proposed in stochastic approximation theory, the concept of iterate averaging reported in \cite{RefWorks:BTPolyak} is an especially appealing and simple way to improve the convergence performance. It was shown in \cite{RefWorks:BTPolyak} that the averaged sequence $\sum_{m=0}^{n-1} \boldsymbol{p}(m)/n$ converges to its limit if step size sequence $a (n)$ decays more slowly than $\mathcal{O} (\frac{1}{n})$ used in the original Robbins-Monro formulation \cite{RefWorks:Robbins2}. This iterate averaging method is optimal in terms of convergence rate. We will take advantage of this appealing technique to improve the convergence performance of SDLA-I.

Using the concept of iterate averaging, we add an averaging operation to the basic recursion (\ref{eq:9}) in SDLA-I, i.e.,
\begin{align}
\boldsymbol{p_j} (n+1) &= \mathcal{P}_{\Phi_j} [\boldsymbol{p_j} (n) + 
a_j(n) \frac{\boldsymbol{\hat{f}_j} (n)}{\boldsymbol{p_j} (n)}], \notag \\
\boldsymbol{\tilde{p}_j} (n+1) &= \frac{1}{n+1} (n \boldsymbol{\tilde{p}_j} (n) +   \boldsymbol{p_j} (n+1) ).
\label{eq:266}
\end{align}
The stochastic learning algorithm with the above modified recursion will be referred to as SDLA-II. It can be shown $(\boldsymbol{\tilde{p}}(n))_{n=0}^{\infty}$ generated by SDLA-II converges to the unique  NE $\boldsymbol{p^*}$ of the stochastic game $\mathcal{G}$ in the mean square sense, i.e., $\lim_{n\rightarrow \infty} \parallel \boldsymbol{p}(n) - \boldsymbol{p^*} \parallel^2 =0$ almost surely, as long as $a_j(n)$ is a suitable decreasing sequence or even fixed step size sequence with small enough value. A detail proof for the convergence of SDLA-II can be carried out by following similar arguments as \cite{RefWorks:BTPolyak} and is thus omitted here. We instead provide an intuitive exposition on why SDLA-II has faster convergence rate than SDLA-I. The idea behind SDLA-II is that we can use larger step size in the basic online recursion for $\boldsymbol{p} (n)$ and the increased noise effects due to larger step size can be smoothed out by the offline averaging recursion for $\boldsymbol{\tilde{p}} (n)$. As a result, SDLA-II converges faster with larger step size and is less likely to get stuck at the first few iterations \cite{RefWorks:Harold}. Indeed, our numerical results demonstrate the convergence rate improvement of SDLA-II over SDLA-I.

A careful reflection on the power allocation trajectory $ (\boldsymbol{p}(n),\boldsymbol{\tilde{p}}(n))_{n=0}^{\infty}$ generated by SDLA-II may reveal a potential handicap in guaranteeing better convergence performance of SDLA-II over SDLA-I. Specifically, with arbitrarily initial starting point $\boldsymbol{p}(0)$ which is quite unlikely near the desired solution of NE $\boldsymbol{p^*}$, it is expected that $\boldsymbol{p}(n)$ moves in an approximately monotonic fashion to the NE $\boldsymbol{p^*}$ at the early stage since the channel power gains of parallel Gaussian interference channels satisfy $\boldsymbol{\Gamma} \succ \boldsymbol{0}$ and thus the  underlying driving force $\boldsymbol{\bar{s}}$ of the recursion of $\boldsymbol{p}(n)$ is strongly monotone. Nevertheless, the noise due to the randomness plays a relatively significant role in the recursion compared to the underlying driving force $\boldsymbol{\bar{s}}$ after sufficient number of iterations. In other words, $\boldsymbol{p}(n)$ starts to hover randomly around the NE $\boldsymbol{p^*}$ when $\boldsymbol{p}(n)$ is near NE $\boldsymbol{p^*}$. Only at this stage can the iterate averaging $\boldsymbol{\tilde{p}}(n)$ be successful since the averaging in this stage can produce a mean solution that is nearer to the NE $\boldsymbol{p^*}$. This implies that the communication pairs should transmit with power level $\boldsymbol{p} (n)$ generated by the basic recursion at the initial stage, and transmit with power level $\boldsymbol{\tilde{p}} (n)$ generated by averaging after SDLA-II has sufficiently converged.

The above reflection does not imply that SDLA-II is of limited use in practical communication systems. Indeed, SDLA-II has few gains in terms of convergence rate over SDLA-I when the initial stage is relatively long compared to the communication period. Nevertheless, the communication time scale can be large in common applications such as video transmission in wireless data networks \cite{RefWorks:vanderSchaar}. Thus, SDLA-II will yield better estimate of NE $\boldsymbol{p^*}$ over SDLA-I in the long run.

\section{Numerical Results}
\label{sec:numerical}

We provide some numerical results in this section for illustration purposes. Simulation parameters are chosen as follows unless specified otherwise. Inspired by \cite{RefWorks:ZQLuo2} and \cite{RefWorks:Setoodeh}, we set both the number of users and number of channels to be $4$. The channel power gains $g^k_{ij}$ are chosen randomly from the intervals $(\bar{g}^k_{ij}(1-\upsilon), \bar{g}^k_{ij}(1+\upsilon) )$ with $\upsilon \in \{10\%, 20\%, 30\%, 40\%, 50\% \}$. Clearly, perturbation parameter $\upsilon$ can serve as an indicator for the time varying rates of parallel Gaussian interference channels. In particular, larger $\upsilon$ implies faster channel varying rate. We further let $\bar{g}^k_{ij} = 15$ if $i=j$ and $0.75$ otherwise. With this choice of simulation parameters, one can verify that $\boldsymbol{\Gamma} \succ \boldsymbol{0}$ almost surely if $\upsilon \in \{10\%, 20\%, 30\% \}$. For clarity, we relax the spectral constraints, i.e., $\bar{p}^k_i = +\infty, \forall i \in \mathcal{N}, \forall k \in \mathcal{K}$. The total power constraint $p_i^{max} = 10*K = 40, \forall i \in \mathcal{N}$. Besides, the background noise level $n_i^k = 0.1/N = 0.025, \forall i \in \mathcal{N}, \forall k \in \mathcal{K}$. We also choose common step size for all users. So we simply write $a_i (n)$ as $a_n$ in this section.

We first compare our proposed SDLA-I with the popular IWFA. We let users using IWFA have the perfect CSI and interference levels at the corresponding transmitters in every power update, while users implementing SDLA-I only have stochastic gradients subject to errors. Due to the limited space, we only show the power evolution of user $1$ on channel $1$ as a function of iteration index in Fig. 1. As expected, even with perfect CSI and interference level, the power evolution generated by IWFA fluctuates significantly. In contrast, users in SDLA-I are more concerned about the long term transmission rates. Consequently, the power evolution only fluctuates mildly after sufficiently long period of learning about the environment. Another interesting observation here is that constant step sizes also lead to convergence of SDLA-I. Indeed, one can show that $(\boldsymbol{p}(n))_{n=0}^{\infty}$ converges to the neighborhood of the unique NE $\boldsymbol{p^*}$ with the choice of sufficiently small constant step sizes. We omit the details due to limited space.

Though both constant step size $\mathcal{O} (1)$ and decreasing step size $\mathcal{O} (\frac{1}{n})$ can lead to the convergence of SDLA-I in numerical experiments, we observe that a tradeoff exists between the convergence rate and exactness of the converged value, which is evaluated by the standard normalized squared error (NSE) defined as 
\begin{align}
NSE (n) = \parallel \boldsymbol{p}(n) - \boldsymbol{p}^* \parallel / \parallel  \boldsymbol{p}^* \parallel.
\label{eq:num1}
\end{align}
The numerical results are shown in Fig. \ref{fig:3}. As shown, decreasing step size $\mathcal{O} (\frac{1}{n})$ has better convergence rate than constant step size $\mathcal{O} (1)$ since $\mathcal{O} (\frac{1}{n})$ goes to $0$ very fast and thus new channel power gain realization has little effect on power update. However, the solution obtained by decreasing step size is not as exact as those by constant step sizes. Nevertheless, an appropriate choice of constant step size is necessary to trade off the convergence rate and exactness of the converged value. Indeed, the convergence rate with $a_n = 0.01$ is very slow as shown in Fig. \ref{fig:3}. Besides, numerical results in Fig. \ref{fig:3} also demonstrate the exponential convergence rate predicted by the continuous time approximation using PDS in section \ref{sec:PDS}.

We show the impact of time-varying rate in parallel Gaussian interference channels on the convergence performance of SDLA-I in Fig. \ref{fig:4}. As described, perturbation parameter $\upsilon$ can be used to model the time varying rate of parallel Gaussian interference channels in our setting. The power evolutions of user $1$ on channel $1$ as a function of iteration index are plotted with different $\upsilon$'s in Fig. \ref{fig:4}. It is shown that the power allocation does not converge when $\upsilon \in \{40\%, 50\%\}$. Indeed, one can verify that $\boldsymbol{\Gamma} \succ \boldsymbol{0}$ can not hold almost surely with $\upsilon \in \{40\%, 50\%\}$. Thus, the convergence of SDLA-I is not guaranteed by Theorem 1. Note that $\boldsymbol{\Gamma} \succ \boldsymbol{0}$ is also required in one way or another in existing distributed power control algorithms including IWFA for deterministic parallel Gaussian interference channels. We in this numerical example also observe the importance of condition $\boldsymbol{\Gamma} \succ \boldsymbol{0}$ for the power control in stochastic parallel Gaussian interference channels.

We next show the performance improvement by iterate averaging in terms of convergence rate. In Fig. \ref{fig:5}, users transmit with power level $\boldsymbol{p} (n)$ under SDLA-I. In contrast, under pure SDLA-II, users transmit with power level $\boldsymbol{\tilde{p}} (n)$ which is generated by averaging $\boldsymbol{p} (n)$. The mixed SDLA-III in Fig. \ref{fig:5} represents a transmission scenario, where users transmit with power level $\boldsymbol{p} (n)$ at the first $100$ iterations, and afterwards transmit with power level $\boldsymbol{\tilde{p}} (n)$ generated by averaging $\boldsymbol{p} (n)$ from the $101$-th iteration. As expected, the iterate averaging $\boldsymbol{\tilde{p}}(n)$ starts to work after $\boldsymbol{p} (n)$ is near to the NE $\boldsymbol{p^*}$.

\section{Conclusion}
\label{sec:conclusion}

In this paper, we investigate the distributed power control problem for stochastic parallel Gaussian interference channels. We formulate the problem in question as a noncoperative stochastic game $\mathcal{G}$. New challenges arise since users in game $\mathcal{G}$ cannot even know their exact utility functions. With these difficulties, we first propose a basic learning algorithm SDLA-I to help users learn the NE in a distributed fashion. The convergence property of SDLA-I is carefully analyzed using stochastic approximation theory. Besides, we provide a continuous time approximation by PDS to appreciate the convergence speed of SDLA-I. Inspired by the recent developments in stochastic approximation theory, we also propose another learning algorithms SDLA-II by including a simple iterate averaging idea into SDLA-I. Numerical results are provided to demonstrate the theoretical results and algorithms. Since existing works only considered deterministic transmission scenarios, our work fills the gap by studying the distributed power control problem in stochastic transmission scenarios.

\appendices

\section{Proof of Proposition 1}
\label{proof:pro1}
\begin{proof}
It is obvious that $\boldsymbol{\Phi}$ is a convex, nonempty, and compact set. Besides, $\bar{R}_j (\boldsymbol{p_{j}}, \boldsymbol{p_{-j}})$ is jointly continuous by assumption. Noting further that $R_j (\boldsymbol{p_{j}}, \boldsymbol{p_{-j}} | \boldsymbol{g})$ is concave with respect to $\boldsymbol{p_{j}}$, we conclude that $\bar{R}_j (\boldsymbol{p_{j}}, \boldsymbol{p_{-j}}) = \mathbb{E}[R_j (\boldsymbol{p_{j}}, \boldsymbol{p_{-j}} | \boldsymbol{G})]$ is also concave with respect to $\boldsymbol{p_{j}}$ since expectation operation preserves concavity. The existence of NE thus follows from standard results in game
theory \cite{RefWorks:Martin}.
\end{proof}

\section{Proof of Proposition 2}
\label{proof:pro2}
\begin{proof}
We can prove this proposition by analyzing the well-known Karush-Kuhn-Tucker (KKT) conditions in optimization theory \cite{RefWorks:Scutari2} \cite{RefWorks:Boyd}. To begin with, note that the projection operation (\ref{eq:9}) is equivalent to the following optimization problem:
\begin{align}
\textrm{minimize} \   & \frac{1}{2}\parallel \boldsymbol{p_j} (n+1) - (\boldsymbol{p_j} (n) + a_j(n) \frac{\boldsymbol{\hat{f}_j}(n)}{\boldsymbol{p_j} (n)}) \parallel^2_2 \notag \\
\textrm{subject to} \  & \sum_{k \in \mathcal{K}} p^k_j (n+1) \leq p_j^{max},\notag \\
& 0 \leq p^k_j (n+1) \leq \bar{p}^k_j, \forall k \in \mathcal{K}. 
\label{eq:13}
\end{align}
This quadratic optimization problem is strictly convex. Therefore, the corresponding solution $\boldsymbol{p_j} (n+1)$ can be obtained from the KKT conditions which are both necessary and sufficient for the optimality \cite{RefWorks:Boyd}. Toward this end, consider the Lagrangian:
\begin{align}
\mathcal{L} (\boldsymbol{p_{j}} (n+1),\lambda_{j}, \boldsymbol{u_{j}}, \boldsymbol{v_{j}}) 
= & \frac{1}{2}\sum_{k \in \mathcal{K}} (p_j^k (n+1) - (p_j^k (n) + a_j(n)\frac{\hat{f}_{j}^k(n)}{p_{j}^k(n)} ))^2 +\lambda_{j} (\sum_{k \in \mathcal{K}} p_{j}^k (n+1)- p_{j}^{max}) \notag \\
 &+ \sum_{k \in \mathcal{K}} u_{j}^k (p_{j}^k (n+1) - \bar{p}^k_j) - \sum_{k \in \mathcal{K}} v_{j}^k p_{j}^k (n+1), 
\label{eq:14}
\end{align}
where $\lambda_{j}$, $\boldsymbol{u_{j}}=[u_{j}^1,..., u_{j}^K]^T$, $\boldsymbol{v_{j}}=[v_{j}^1,..., v_{j}^K]^T$ are the associated Lagrangian multipliers. Then the KKT conditions are given by 
\begin{align}
p_j^k (n+1) - (p_j^k (n) + a_j(n)\frac{\hat{f}_{j}^k(n)}{p_{j}^k(n)} ) + \lambda_{j} + u_{j}^k - v_{j}^k &= 0, \forall k \in \mathcal{K} \notag \\
u_{j}^k \geq 0, \ p_{j}^k (n+1) \leq \bar{p}_{j}^k,\ u_{j}^k (p_{j}^k (n+1) -\bar{p}_{j}^k) &=0, \forall k \in \mathcal{K} \notag \\
v_{j}^k \geq 0,\ p_{j}^k (n+1)\geq 0,\ v_{j}^k p_{j}^k (n+1) &=0, \forall k \in \mathcal{K} \notag \\
 \lambda_{j} \geq 0, \sum_{k \in \mathcal{K}} p_{j}^k (n+1) \leq p_{j}^{max}, \lambda_{j} (\sum_{k \in \mathcal{K}} p_{j}^k (n+1) - p_{j}^{max})&= 0. 
\label{eq:15}
\end{align}
Now for any $k \in \mathcal{K}$, we observe that if $p_j^k (n+1) = 0$, we have $u_{j}^k =0$ by complementary slackness condition. Furthermore, we have
$-v_j^k=p_j^k (n) + a_j(n)\frac{\hat{f}_{j}^k(n)}{p_{j}^k(n)}  - \lambda_j \leq 0$.
By a similar argument, we can obtain that $p_j^k (n+1) = p_j^k (n) + a_j(n)\frac{\hat{f}_{j}^k(n)}{p_{j}^k(n)}  - \lambda_j$ if $0 < p_j^k (n+1) < \bar{p}_{j}^k$, and $p_j^k (n) + a_j(n)\frac{\hat{f}_{j}^k(n)}{p_{j}^k(n)}  - \lambda_j \geq \bar{p}_{j}^k$ if $p_j^k (n+1) = \bar{p}_{j}^k$. This completes the proof.
\end{proof}

\section{Proof of Lemma 1}
\label{proof:lem1}
\begin{proof}
(i). We know that if $\boldsymbol{p^*} \in \Phi$ is an NE, then for any $a_j >0$ \cite{RefWorks:Bertsekas}
\begin{align}
\boldsymbol{p_j^*} = \mathcal{P}_{\Phi_j} [\boldsymbol{p_j^*} + a_j \nabla_{\boldsymbol{p_j}} \mathbb{E}[R_j (\boldsymbol{p_j^*}, \boldsymbol{p_{-j}^*} | \boldsymbol{G})] ],  \forall j \in \mathcal{N}.
\end{align}
The result follows if the interchange of mathematical expectations and gradient signs is justified. Recall that the realization $\boldsymbol{g}$ is bounded by assumption. Then it is straightforward to verify $\parallel \nabla_{\boldsymbol{p_j}} R_j (\boldsymbol{p_{j}}, \boldsymbol{p_{-j}} | \boldsymbol{g}) \parallel$ is also bounded. Thus, $\nabla_{\boldsymbol{p_j}} \mathbb{E}[R_j (\boldsymbol{p_j^*}, \boldsymbol{p_{-j}^*} | \boldsymbol{G})] = \mathbb{E}[\nabla_{\boldsymbol{p_j}} R_j (\boldsymbol{p_j^*}, \boldsymbol{p_{-j}^*} | \boldsymbol{G})]$ \cite{RefWorks:RTRockafellar}.

(ii). This is a well-known result on the nonexpansive property of projection, the proof of which can be found in, e.g., \cite{RefWorks:Bertsekas}.

(iii). Since $\mathcal{P}_{\Phi} (\boldsymbol{p})$ minimizes $\frac{1}{2} \parallel \boldsymbol{\bar{p}} - \boldsymbol{p}\parallel_2^2$ over all $\boldsymbol{\bar{p}} \in \Phi$, we have
\begin{align}
(\boldsymbol{\bar{p}} - \mathcal{P}_{\Phi} (\boldsymbol{p}))^T(\mathcal{P}_{\Phi} (\boldsymbol{p}) - \boldsymbol{p}) \geq 0, \forall \boldsymbol{p} \in \mathbb{R}^{NK},
\end{align}
by optimality condition \cite{RefWorks:Bertsekas}. Noting another obvious fact:
\begin{align}
(\mathcal{P}_{\Phi} (\boldsymbol{p}) - \boldsymbol{p})^T (\mathcal{P}_{\Phi} (\boldsymbol{p}) - \boldsymbol{p}) \geq 0, \forall \boldsymbol{p} \in \mathbb{R}^{NK},
\end{align}
we conclude that $(\boldsymbol{\bar{p}} - \boldsymbol{p})^T(\mathcal{P}_{\Phi} (\boldsymbol{p}) - \boldsymbol{p} ) \geq 0$ for any $\boldsymbol{p} \in \mathbb{R}^{NK}$ and $\boldsymbol{\bar{p}} \in \Phi$.
\end{proof}

\section{Proof of Lemma 2}
\label{proof:lem2}
\begin{proof}
Following Proposition 2 in \cite{RefWorks:JSPang}, if $\boldsymbol{\Gamma} \succ \boldsymbol{0}$ under given $\boldsymbol{g}$, then
\begin{equation}
(\boldsymbol{s} (\boldsymbol{q} | \boldsymbol{g} ) -  \boldsymbol{s} (\boldsymbol{p} | \boldsymbol{g} )  )^T (\boldsymbol{p} - \boldsymbol{q}) \geq \tau(\boldsymbol{s}) \parallel \boldsymbol{p} - \boldsymbol{q} \parallel_2^2, \forall \boldsymbol{p}, \boldsymbol{q} \in \Phi,
\label{eq:lem2}
\end{equation}
with $\tau(\boldsymbol{s})$ specified in (\ref{eq:lem11}). That is, $\boldsymbol{s} (\cdot | \boldsymbol{g} ) $ is strongly monotone on $\Phi$. The uniqueness of NE $\boldsymbol{p^*}(\boldsymbol{g}) \in \Phi$ follows (see, e.g., \cite{RefWorks:Francisco}).

Furthermore, by the equivalence of standard NE problem and variational inequality (VI)\footnote{Given a set $K \subseteq \mathbb{R}^n$ and a mapping $F: K \rightarrow \mathbb{R}^n$, the variational inequality $VI(K,F)$ is to find a vector $x \in K$ such that $(y-x)^T F(x)\geq 0, \forall y \in K$ \cite{RefWorks:Francisco}.}, we have
\begin{align}
-\boldsymbol{s} (\boldsymbol{p^*} | \boldsymbol{g} )^T (\boldsymbol{p}-\boldsymbol{p^*}) \geq 0, \forall \boldsymbol{p} \in \Phi.
\label{eq:lem3}
\end{align}
Substituting $\boldsymbol{p^*}$ for $\boldsymbol{q}$ in (\ref{eq:lem2}), we obtain
\begin{equation}
(\boldsymbol{s} (\boldsymbol{p^*} | \boldsymbol{g} ) -  \boldsymbol{s} (\boldsymbol{p} | \boldsymbol{g} )  )^T (\boldsymbol{p} - \boldsymbol{p^*}) \geq \tau(\boldsymbol{s}) \parallel \boldsymbol{p} - \boldsymbol{p^*} \parallel^2, \forall \boldsymbol{p} \in \Phi.
\label{eq:lem4}
\end{equation}
Thus, the desired inequality (\ref{eq:lem1}) immediately follows from (\ref{eq:lem3}) and (\ref{eq:lem4}). This completes the proof.
\end{proof}

\section{Proof of Theorem 1}
\label{proof:thm1}
\begin{proof}
We first derive a recursion inequality characterizing the relationship between $\parallel \boldsymbol{p} (n) - \boldsymbol{p^*} \parallel$ and $\parallel \boldsymbol{p} (n+1) - \boldsymbol{p^*} \parallel$ in the following lemma. 
\begin{lem} 
The sequence $(\boldsymbol{p} (n))_{n=0}^{\infty}$ generated by iteration (\ref{eq:17}) satisfies
\begin{align}
\parallel \boldsymbol{p} (n+1) - \boldsymbol{p^*}   \parallel^2 \leq & \parallel \boldsymbol{p} (n) - \boldsymbol{p^*}   \parallel^2 +  5C^2  a^2 (n) \notag \\
&+ 2\sum_{i\in \mathcal{N}}a_i(n)\epsilon_i (n) - 2\sum_{i\in \mathcal{N}}a_i(n)\boldsymbol{s_i}(n+1)^T (\boldsymbol{p^*_{i}}-\boldsymbol{p_{i}}(n) ), 
\label{eq:22}
\end{align}
where $\boldsymbol{p^*} \in \Phi$ is any NE, $C$ is some large enough constant, and $a(n) = (\sum_{i \in \mathcal{N}} a^2_i (n))^{\frac{1}{2}}$.
\end{lem}

\begin{proof}
This proof is inspired by constructions from \cite{RefWorks:SDFlam} \cite{RefWorks:AlberYaI}. Consider a fixed trajectory $(\boldsymbol{g}(n))_{n=0}^{\infty}$. Recall that $\boldsymbol{q}(n) = \boldsymbol{p}(n) + D(n) \frac{\boldsymbol{\hat{f}}(n)}{\boldsymbol{p}(n)}$ and $\boldsymbol{p} (n+1) = \mathcal{P}_{\Phi} [\boldsymbol{q}(n)]$. We first have
\begin{align}
\parallel \boldsymbol{p} (n+1) - \boldsymbol{p} (n)   \parallel \leq & \parallel \boldsymbol{q}(n)  - \boldsymbol{p} (n)   \parallel = \parallel \boldsymbol{p}(n) + D(n) \frac{\boldsymbol{\hat{f}}(n)}{\boldsymbol{p}(n)}  - \boldsymbol{p} (n) \parallel \notag \\
= &\parallel D(n) \boldsymbol{\hat{s}}(n) \parallel = (\sum_{i \in \mathcal{N}} a_i^2 (n) \parallel \boldsymbol{\hat{s}_i}(n) \parallel^2)^{\frac{1}{2}} < C(\sum_{i \in \mathcal{N}} a^2_i (n))^{\frac{1}{2}} =C a (n), 
\label{eq:20}
\end{align}
where the first inequality follows from Lemma 1(ii) and $C$ is some large enough constant. The existence of $C$ is guaranteed by the boundedness of $\boldsymbol{\hat{s}}$. We proceed by deriving that
\begin{align}
&C^2 a^2 (n) + \parallel \boldsymbol{p} (n) - \boldsymbol{p^*}   \parallel^2 - \parallel \boldsymbol{p} (n+1) - \boldsymbol{p^*}   \parallel^2 \notag \\
\geq & \parallel \boldsymbol{p} (n+1) - \boldsymbol{p} (n)   \parallel^2 + \parallel \boldsymbol{p} (n) - \boldsymbol{p^*}   \parallel^2 - \parallel \boldsymbol{p} (n+1) - \boldsymbol{p^*}   \parallel^2 \notag \\
= & 2 (\boldsymbol{p} (n+1) - \boldsymbol{p} (n))^T(\boldsymbol{p^*} - \boldsymbol{p} (n)) \notag \\
= & 2 (\boldsymbol{p} (n+1) - \boldsymbol{q}(n) )^T(\boldsymbol{p^*} - \boldsymbol{p} (n)) + 2 (D(n)\boldsymbol{\hat{s}}(n))^T(\boldsymbol{p^*} - \boldsymbol{p} (n)) \notag \\
= & 2 (\boldsymbol{p} (n+1) - \boldsymbol{q}(n) )^T(\boldsymbol{p^*} - \boldsymbol{q} (n))  + 2 (\boldsymbol{p} (n+1) - \boldsymbol{q}(n) )^T(\boldsymbol{q}(n) - \boldsymbol{p} (n)) + 2 (D(n)\boldsymbol{\hat{s}}(n))^T(\boldsymbol{p^*} - \boldsymbol{p} (n)) \notag \\
\geq & 2 (\boldsymbol{p} (n+1) - \boldsymbol{q}(n) )^T(\boldsymbol{q}(n) - \boldsymbol{p} (n)) + 2 (D(n)\boldsymbol{\hat{s}}(n))^T(\boldsymbol{p^*} - \boldsymbol{p} (n)) \notag \\
= & 2 (\boldsymbol{p} (n+1) - \boldsymbol{p}(n) )^T(\boldsymbol{q}(n) - \boldsymbol{p} (n))  + 2 (\boldsymbol{p} (n) - \boldsymbol{q}(n) )^T(\boldsymbol{q}(n) - \boldsymbol{p} (n)) + 2 (D(n)\boldsymbol{\hat{s}}(n))^T(\boldsymbol{p^*} - \boldsymbol{p} (n)) \notag \\
\geq & -2 \parallel \boldsymbol{p} (n+1) - \boldsymbol{p}(n) \parallel \parallel \boldsymbol{q}(n) - \boldsymbol{p} (n) \parallel -2 \parallel \boldsymbol{p} (n) - \boldsymbol{q}(n) \parallel^2 + 2 (D(n)\boldsymbol{\hat{s}}(n))^T(\boldsymbol{p^*} - \boldsymbol{p} (n)) \notag \\
\geq & -4 C^2 a^2 (n) + 2 (D(n)\boldsymbol{\hat{s}}(n))^T(\boldsymbol{p^*} - \boldsymbol{p} (n))
\label{eq:19}
\end{align}
where the first inequality follows from (\ref{eq:20}), the second inequality follows from Lemma 1(iii), and the last inequality also follows from (\ref{eq:20}).

Rearranging terms in (\ref{eq:19}), we proceed as follows:
\begin{align}
\parallel \boldsymbol{p} (n+1) - \boldsymbol{p^*}   \parallel^2 \leq& \parallel \boldsymbol{p} (n) - \boldsymbol{p^*}   \parallel^2 + 5C^2  a^2 (n) - 2 (D(n)\boldsymbol{\hat{s}}(n))^T(\boldsymbol{p^*} - \boldsymbol{p} (n)) \notag \\
=& \parallel \boldsymbol{p} (n) - \boldsymbol{p^*}   \parallel^2 +  5C^2   a^2 (n) - 2 \sum_{i\in \mathcal{N}}(a_i(n)\boldsymbol{\hat{s}_i}(n))^T(\boldsymbol{p_i^*} - \boldsymbol{p_i} (n)) \notag \\
\leq& \parallel \boldsymbol{p} (n) - \boldsymbol{p^*}   \parallel^2 +  5C^2   a^2 (n) + 2\sum_{i\in \mathcal{N}}a_i(n)\epsilon_i (n)\notag \\
&+2\sum_{i\in \mathcal{N}}a_i(n)(R_i (\boldsymbol{p_{i}}(n), \boldsymbol{p_{-i}}(n) | \boldsymbol{g}(n+1)) - R_i (\boldsymbol{p^*_{i}}, \boldsymbol{p_{-i}}(n) | \boldsymbol{g} (n+1)) ) \notag \\
\leq & \parallel \boldsymbol{p} (n) - \boldsymbol{p^*}   \parallel^2 +  5C^2   a^2 (n) + 2\sum_{i\in \mathcal{N}}a_i(n)\epsilon_i (n) + 2\sum_{i\in \mathcal{N}}a_i(n)\boldsymbol{s_i}(n+1)^T (\boldsymbol{p_{i}}(n) - \boldsymbol{p^*_{i}})
\label{eq:21}
\end{align}
where the second inequality follows from (\ref{eq:10}), and the last inequality follows from the concavity of $R_i (\cdot, \cdot | \boldsymbol{g})$ with respect to the first argument. This completes the proof.
\end{proof}

We further need the following well-known lemma in stochastic approximation theory \cite{RefWorks:Robbins}.

\begin{lem}
Let $\{ \mathcal{F}_n \}$ be an increasing sequence of $\sigma$-algebras and $e_n,\alpha_n,\beta_n,\eta_n$ be finite, nonnegative, $\mathcal{F}_n$-measurable random variables. If it holds almost surely that $\sum_{n=0}^{\infty} \alpha_n < \infty, \sum_{n=0}^{\infty} \beta_n < \infty$, and
\begin{align}
\mathbb{E} (e_{n+1} | \mathcal{F}_n) \leq (1+\alpha_n) e_n + \beta_n - \eta_n,
\label{eq:166}
\end{align}
then $(e_{n})_{n=0}^{\infty}$ converges and $\sum_{n=0}^{\infty} \eta_n < \infty$ almost surely. 
\end{lem}

Now taking the conditional expectation $\mathbb{E} [\cdot| \mathcal{F}_n ]$ of both sides in (\ref{eq:22}) yields
\begin{align}
&\mathbb{E} [\parallel \boldsymbol{p} (n+1) - \boldsymbol{p^*}   \parallel^2 | \mathcal{F}_n ] \notag  \\
\leq & \mathbb{E} [\parallel \boldsymbol{p} (n) - \boldsymbol{p^*}   \parallel^2 | \mathcal{F}_n ]  +\mathbb{E} [5 C^2 a^2 (n) + 2\sum_{i\in \mathcal{N}}a_i(n)\epsilon_i (n) | \mathcal{F}_n ] -\mathbb{E} [2\sum_{i\in \mathcal{N}}a_i(n)\boldsymbol{s_i}(n+1)^T (\boldsymbol{p^*_{i}}-\boldsymbol{p_{i}}(n) )| \mathcal{F}_n ] \notag  \\
= & \parallel \boldsymbol{p} (n) - \boldsymbol{p^*}   \parallel^2 +\mathbb{E} [5 C^2 a^2 (n) + 2\sum_{i\in \mathcal{N}}a_i(n)\epsilon_i (n) | \mathcal{F}_n ] -\mathbb{E} [2\sum_{i\in \mathcal{N}}a_i(n)\boldsymbol{s_i}(n+1)^T (\boldsymbol{p^*_{i}}-\boldsymbol{p_{i}}(n) )| \mathcal{F}_n ], 
\label{eq:23}
\end{align}
We see (\ref{eq:166}) is satisfied by substituting $e_n = \parallel \boldsymbol{p} (n) - \boldsymbol{p^*}   \parallel^2$, $\alpha_n = 0$, $\beta_n = \mathbb{E} [ 5 C^2 a^2 (n) + 2\sum_{i\in \mathcal{N}}a_i(n)\epsilon_i (n)  | \mathcal{F}_n ]$, and $\eta_n = \mathbb{E} [ 2\sum_{i\in \mathcal{N}}a_i(n)\boldsymbol{s_i}(n+1)^T (\boldsymbol{p^*_{i}}-\boldsymbol{p_{i}}(n) )  | \mathcal{F}_n ]$. 

Clearly, $e_n$, $\alpha_n$, and $\beta_n$ are finite, nonnegative, $\mathcal{F}_n$-measurable, and $\sum_{n=0}^{\infty} \alpha_n =0 < \infty$. $\sum_{n=0}^{\infty} \beta_n < \infty$ follows from assumption (ii) and (iii) in Theorem 1. $\eta_n$ is also obviously finite, and $\mathcal{F}_n$-measurable. The nonnegativeness of $\eta_n$ follows from Lemma 2 and assumption (i) in Theorem 1. Thus, all the conditions in Lemma 5 are satisfied. We conclude that $e_n = \parallel \boldsymbol{p} (n) - \boldsymbol{p^*}   \parallel^2$ converges almost surely, and that 
\begin{align}
\sum_{n=0}^{\infty} \eta_n = \sum_{n=0}^{\infty} \mathbb{E} [ 2\sum_{i\in \mathcal{N}}a_i(n)\boldsymbol{s_i}(n+1)^T (\boldsymbol{p^*_{i}}-\boldsymbol{p_{i}}(n) )  | \mathcal{F}_n ] 
\label{eq:233}
\end{align}
is finite almost surely.

We still need to show $e_n = \parallel \boldsymbol{p} (n) - \boldsymbol{p^*}   \parallel^2$ converges to $0$ almost surely. If this is not true, the event $A = \{w: \lim_{n \rightarrow \infty} e_n (w) = e(w) > 0 \}$ has nonzero probability where $w$ is a trajectory on the associated probability space. Then for any $w \in A$, there exists a large enough $N(w)$ such that 
\begin{align}
\sum_{n=N(w)}^{\infty}  2\sum_{i\in \mathcal{N}}a_i(n)\boldsymbol{s_i}(n+1)^T (\boldsymbol{p^*_{i}}-\boldsymbol{p_{i}}(n) ) &\geq 2 \sum_{n=N(w)}^{\infty} \min_{i\in \mathcal{N}} a_i (n)  \min_{\boldsymbol{s}} \tau(\boldsymbol{s}) \parallel \boldsymbol{p^*} - \boldsymbol{p}(n) \parallel^2 \notag \\
&\geq 2 \sum_{n=N(w)}^{\infty} \min_{i\in \mathcal{N}} a_i (n) \min_{\boldsymbol{s}} \tau(\boldsymbol{s}) \  e(w) = +\infty
\label{eq:24}
\end{align}
where the first inequality follows from Lemma 2 and the last inequality follows from assumption (ii) in Theorem 1. Since (\ref{eq:24}) happens with nonzero probability, the random sum $\sum_{n=0}^{\infty} \eta_n$ in (\ref{eq:233}) cannot be finite almost surely, resulting in a contradiction. Hence, we conclude that $(\boldsymbol{p}(n))_{n=0}^{\infty}$ converges to $\boldsymbol{p^*}$ almost surely. This completes the proof.
\end{proof}

\section{Proof of Lemma 3}
\label{proof:lem3}
\begin{proof}
Note that we assume that $\boldsymbol{G}$ is bounded almost surely. Given bounded realization $\boldsymbol{g}$, it is straightforward to verify that $\boldsymbol{s}(\boldsymbol{p} | \boldsymbol{g})$ where $\boldsymbol{s_j} (\boldsymbol{p} | \boldsymbol{g}) =  \nabla_{\boldsymbol{p_j}}  R_j(\boldsymbol{p_{j}}, \boldsymbol{p_{-j}} | \boldsymbol{g}) $ has bounded derivative and thus Lipschitz continuous. It follows that $\boldsymbol{\bar{s}}(\boldsymbol{p})$ is Lipschitz continuous almost surely. 
\end{proof}

\section{Proof of Theorem 2}
\label{proof:thm2}

\begin{proof}
The proof essentially follows the same arguments as the proof of Theorem 1. Specifically, we first observe that 
\begin{align}
&\mathbb{E} [\parallel \boldsymbol{p} (n+1) - \boldsymbol{p^*}   \parallel^2 | \mathcal{F}_n ]  \notag  \\
=&\mathbb{E} [\parallel \mathcal{P}_{\Phi}( \boldsymbol{p} (n) + D(n) \boldsymbol{\hat{s}}(n) ) - \mathcal{P}_{\Phi} (\boldsymbol{p^*} + D(n)\boldsymbol{\bar{s}}(\boldsymbol{p^*}) )   \parallel^2 | \mathcal{F}_n ]   \notag  \\
\leq & \mathbb{E} [\parallel  \boldsymbol{p} (n) - \boldsymbol{p^*} + D(n) (\boldsymbol{\hat{s}}(n)   - \boldsymbol{\bar{s}}(\boldsymbol{p^*}) )    \parallel^2 | \mathcal{F}_n ] \notag  \\
=&\mathbb{E} [\parallel  \boldsymbol{p} (n) - \boldsymbol{p^*} + D(n) (\boldsymbol{\bar{s}}(\boldsymbol{p}(n)) + \boldsymbol{\theta}(n)   - \boldsymbol{\bar{s}}(\boldsymbol{p^*}) )    \parallel^2 | \mathcal{F}_n ] \notag  \\
=&\parallel \boldsymbol{p} (n) - \boldsymbol{p^*} \parallel^2 +\sum_{i \in \mathcal{N}} a^2_i(n)  \parallel  \boldsymbol{\bar{s}_i}(\boldsymbol{p}(n)) - \boldsymbol{\bar{s}_i}(\boldsymbol{p^*})  \parallel^2 +2 \sum_{i \in \mathcal{N}} a_i(n) ( \boldsymbol{\bar{s}_i}(\boldsymbol{p}(n)) - \boldsymbol{\bar{s}_i}(\boldsymbol{p^*}) )^T (\boldsymbol{p_i} (n) - \boldsymbol{p_i^*}) \notag  \\
&+ \mathbb{E} [ \parallel D(n) \boldsymbol{\theta}(n) \parallel^2 | \mathcal{F}_n] + 2( \boldsymbol{p} (n) - \boldsymbol{p^*} )^T \mathbb{E} [ D(n) \boldsymbol{\theta}(n)  | \mathcal{F}_n] + 2(\boldsymbol{\bar{s}}(\boldsymbol{p}(n)) - \boldsymbol{\bar{s}}(\boldsymbol{p^*}))^T \mathbb{E} [ D^2(n) \boldsymbol{\theta}(n)  | \mathcal{F}_n]\notag  \\
=&\parallel \boldsymbol{p} (n) - \boldsymbol{p^*} \parallel^2 +  \mathbb{E} [ \parallel D(n) \boldsymbol{\theta}(n) \parallel^2 | \mathcal{F}_n] +\sum_{i \in \mathcal{N}}     a^2_i(n) \parallel  \boldsymbol{\bar{s}_i}(\boldsymbol{p}(n)) - \boldsymbol{\bar{s}_i}(\boldsymbol{p^*})  \parallel^2  \notag  \\
&+2 \sum_{i \in \mathcal{N}} a_i(n)  ( \boldsymbol{\bar{s}_i}(\boldsymbol{p}(n)) - \boldsymbol{\bar{s}_i}(\boldsymbol{p^*}) )^T (\boldsymbol{p_i} (n) - \boldsymbol{p_i^*}) \notag  \\
\leq &\parallel \boldsymbol{p} (n) - \boldsymbol{p^*} \parallel^2 +  \mathbb{E} [ \parallel D(n) \boldsymbol{\theta}(n) \parallel^2 | \mathcal{F}_n] + \max_{i \in \mathcal{N}} a^2_i(n) \sum_{i \in \mathcal{N}}  \parallel  \boldsymbol{\bar{s}_i}(\boldsymbol{p}(n)) - \boldsymbol{\bar{s}_i}(\boldsymbol{p^*})  \parallel^2  \notag  \\
&+2 \sum_{i \in \mathcal{N}} a_i(n) ( \boldsymbol{\bar{s}_i}(\boldsymbol{p}(n)) - \boldsymbol{\bar{s}_i}(\boldsymbol{p^*}) )^T (\boldsymbol{p_i} (n) - \boldsymbol{p_i^*}) \notag  \\
\leq &\parallel \boldsymbol{p} (n) - \boldsymbol{p^*} \parallel^2 +  \mathbb{E} [ \parallel D(n) \boldsymbol{\theta}(n) \parallel^2 | \mathcal{F}_n] + L^2 \max_{i \in \mathcal{N}} a^2_i(n)  \sum_{i \in \mathcal{N}}  \parallel  (\boldsymbol{p_i}(n) - \boldsymbol{p_i^*})  \parallel^2 \notag  \\
&- 2 \min_{i\in \mathcal{N}} a_{i} (n) \tau(\boldsymbol{\bar{s}})   \parallel \boldsymbol{p} (n) - \boldsymbol{p^*} \parallel^2 \notag  \\
=&\parallel \boldsymbol{p} (n) - \boldsymbol{p^*} \parallel^2 + \sum_{i \in \mathcal{N}}a^2_i (n)   \mathbb{E}[\parallel \boldsymbol{\theta_i}(n)  \parallel^2 | \mathcal{F}_n] - (2 \tau(\boldsymbol{\bar{s}}) \min_{i\in \mathcal{N}} a_{i} (n)  - L^2 \max_{i \in \mathcal{N}} a^2_i(n)) \parallel \boldsymbol{p} (n) - \boldsymbol{p^*} \parallel^2. 
\label{eq:25}
\end{align}
Here the first equality follows from (\ref{eq:16}). The first inequality follows from (\ref{eq:1616}). The fourth equality follows from assumption $\mathbb{E} [\boldsymbol{\theta}(n) | \mathcal{F}_n] = \boldsymbol{0}$. The last inequality follows from Lemma 3, assumption (i) in Theorem 2, and Lemma 2.

Substitute $e_n = \parallel \boldsymbol{p} (n) - \boldsymbol{p^*}   \parallel^2$, $\alpha_n = 0$, $\beta_n = \sum_{i \in \mathcal{N}}a^2_i (n)   \mathbb{E}[\parallel \boldsymbol{\theta_i}(n)  \parallel^2 | \mathcal{F}_n]$, and $\eta_n = (2 \tau(\boldsymbol{\bar{s}}) \min_{i\in \mathcal{N}} \\ a_{i} (n)   - L^2 \max_{i \in \mathcal{N}} a^2_i(n)) \parallel \boldsymbol{p} (n) - \boldsymbol{p^*} \parallel^2$. It is straightforward to verify that all the assumptions in Lemma 5 are satisfied. We conclude that $e_n = \parallel \boldsymbol{p} (n) - \boldsymbol{p^*}   \parallel^2$ converges almost surely, and that 
\begin{align}
\sum_{n=0}^{\infty} \eta_n = \sum_{n=0}^{\infty} (2 \tau(\boldsymbol{\bar{s}}) \min_{i\in \mathcal{N}} a_{i} (n)  - L^2 \max_{i \in \mathcal{N}} a^2_i(n)) \parallel \boldsymbol{p} (n) - \boldsymbol{p^*} \parallel^2
\label{eq:26}
\end{align}
is finite almost surely.

We further claim that $e_n = \parallel \boldsymbol{p} (n) - \boldsymbol{p^*}   \parallel^2$ converges to $0$ almost surely. Observe that $2 \tau(\boldsymbol{\bar{s}}) \min_{i\in \mathcal{N}} a_{i} (n) \\ - L^2 \max_{i \in \mathcal{N}} a^2_i(n)$ is bounded away from $0$ by assumption (ii), this claim holds by following a similar argument by contrapositive as that of the proof for Theorem 1.
\end{proof}

\section{Proof of Proposition 3}
\label{proof:pro3}
\begin{proof}
(i). Note that $\boldsymbol{\bar{s}} (p)$ is Lipschitz continuous by Lemma 3. Hence $PDS(\boldsymbol{\bar{s}},\Phi)$ is well posed and the results follow from Theorem 2.5 in \cite{Refworks:Anna}.

(ii). The equivalence of NE in game $\mathcal{G}$ and the set of stationary points in PDS can be shown by observing that $\Phi$ is convex polyhedron by following \cite{Refworks:Anna}. We provide a sketch of the proof here for completeness. Define a variational inequality problem $VI(\boldsymbol{\bar{s}}, \Phi)$, the aim of which is to find a vector $\boldsymbol{p^*}$ such that 
\begin{equation}
(\boldsymbol{p}-\boldsymbol{p^*})^T \boldsymbol{\bar{s}(p^*)} \leq 0, \forall \boldsymbol{p} \in \Phi.
\label{eq:PDS4}
\end{equation} 
It is known that $\boldsymbol{p^*}$ is a solution to $VI(\boldsymbol{\bar{s}}, \Phi)$ if and only if it is an NE of the game $\mathcal{G}$ (see, e.g., \cite{RefWorks:Francisco}). Noting that $\Phi$ is convex polyhedron, the stationary points of $PDS(\boldsymbol{\bar{s}},\Phi)$ coincide with the solutions of $VI(\boldsymbol{\bar{s}}, \Phi)$ by Theorem 2.4 in \cite{Refworks:Anna}. 

(iii). Recall the condition that $\boldsymbol{\Gamma} \succ \boldsymbol{0}$ holds almost surely implies the strongly monotonicity of $\boldsymbol{\bar{s}} (\boldsymbol{p})$. Then the uniqueness and globally exponential stability follow from Theorem 3.7 in \cite{Refworks:Anna}. Indeed, we can associate a Liapunov function $\parallel \boldsymbol{p} - \boldsymbol{p^*} \parallel$ for $PDS(\boldsymbol{\bar{s}},\Phi)$ to obtain the stability result.
\end{proof}

%





\bibliographystyle{IEEEtran}
%

\bibliography{IEEEabrv,first}	

\newpage

\begin{figure}
\centering
\begin{minipage}[b]{0.8\textwidth}
\includegraphics[width=0.8\textwidth]{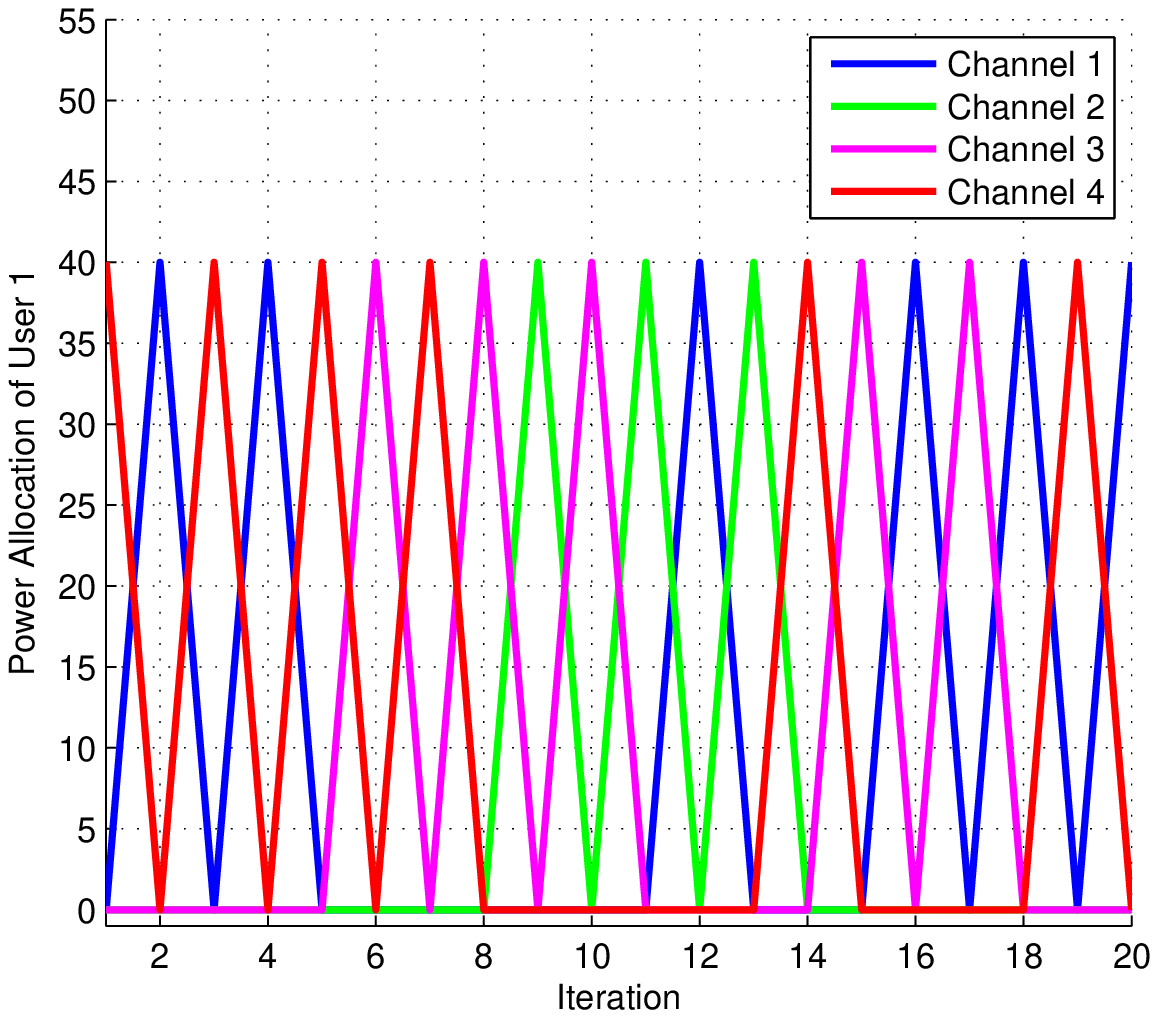}
\includegraphics[width=0.8\textwidth]{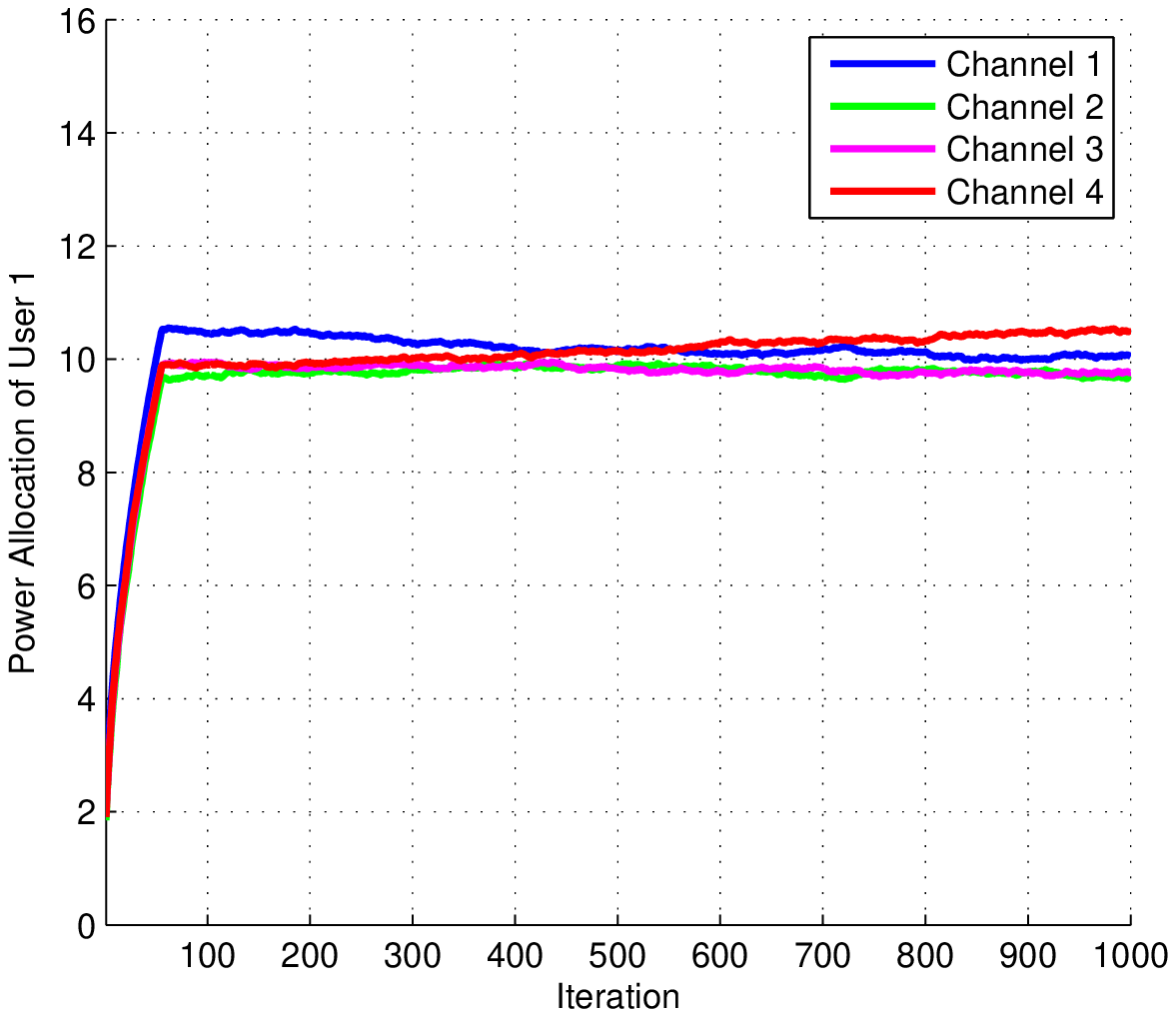} 
\end{minipage}
\label{fig:2}
\caption{Comparison of IWFA and SDLA-I: In SDLA-I, $\upsilon = 20\%$ and $a_n = 0.5$.}
\end{figure}

\setlength{\floatsep} {10pt plus 3pt minus 2pt}
\setlength{\textfloatsep} {12pt plus 2pt minus 2pt}
\setlength{\abovecaptionskip}{0pt}
\setlength{\belowcaptionskip}{0pt}
\begin{figure}
\centering
\includegraphics[width=14cm]{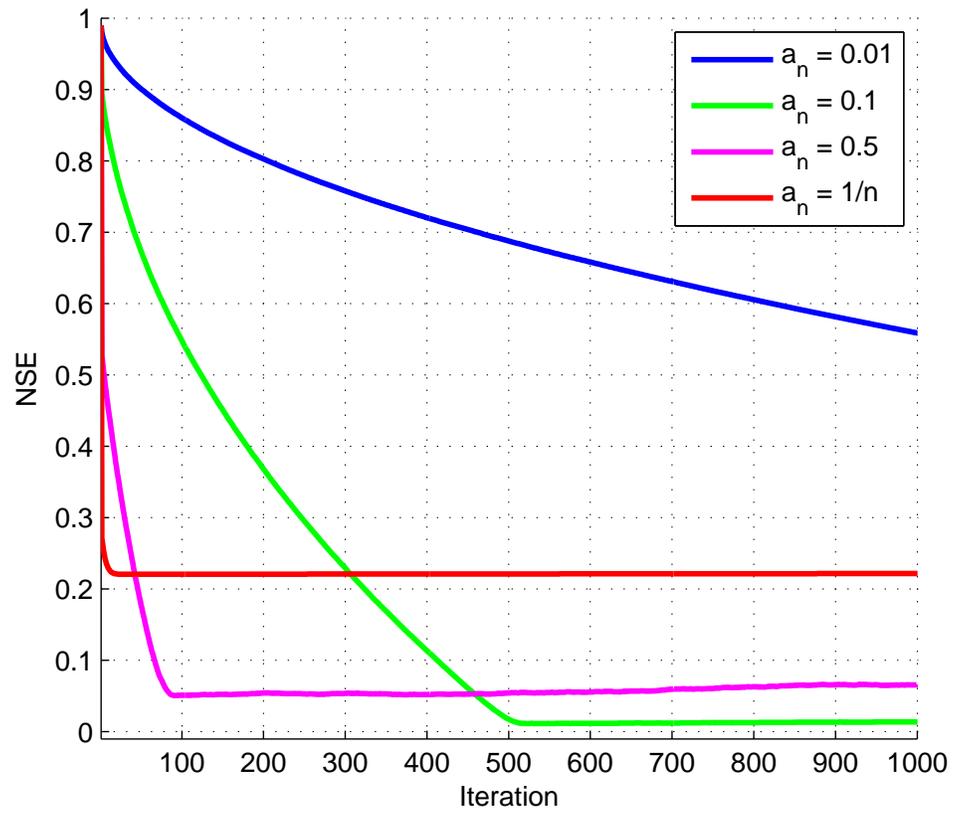}
\caption{Impact of Step Sizes:  $\upsilon = 20\%$.}
\label{fig:3}
\end{figure}

\setlength{\floatsep} {10pt plus 3pt minus 2pt}
\setlength{\textfloatsep} {12pt plus 2pt minus 2pt}
\setlength{\abovecaptionskip}{0pt}
\setlength{\belowcaptionskip}{0pt}
\begin{figure}
\centering
\includegraphics[width=14cm]{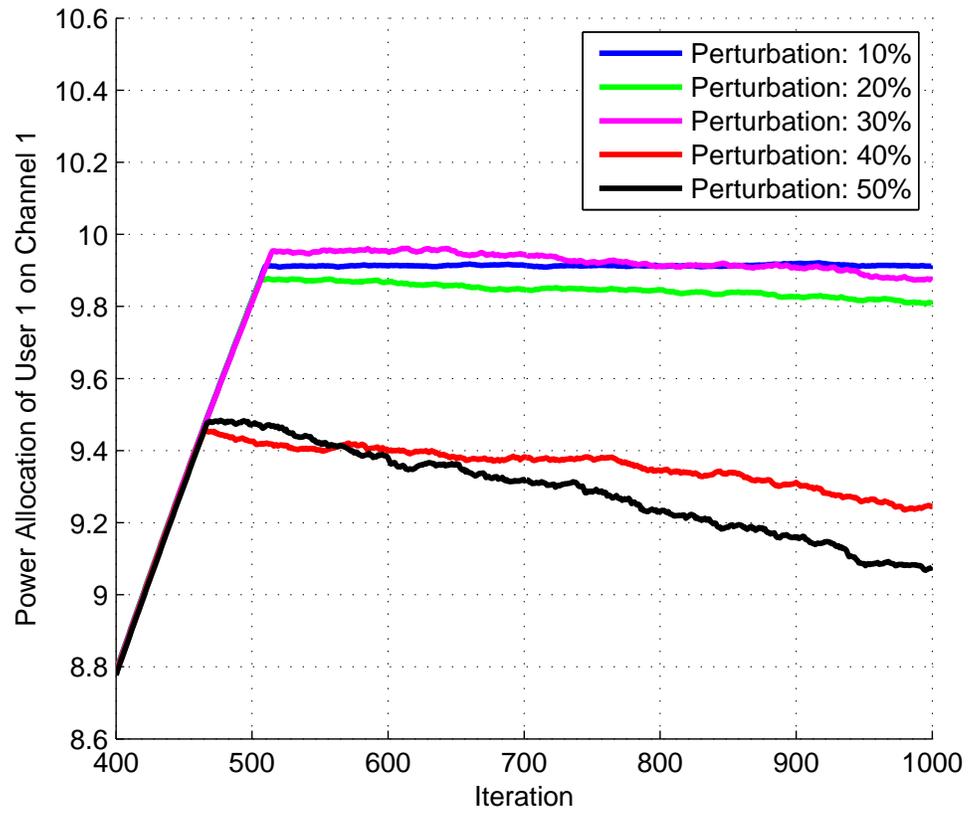}
\caption{Impact of Time-Varying Rate: $a_n = 0.1$.}
\label{fig:4}
\end{figure}

\setlength{\floatsep} {10pt plus 3pt minus 2pt}
\setlength{\textfloatsep} {12pt plus 2pt minus 2pt}
\setlength{\abovecaptionskip}{0pt}
\setlength{\belowcaptionskip}{0pt}
\begin{figure}
\centering
\includegraphics[width=14cm]{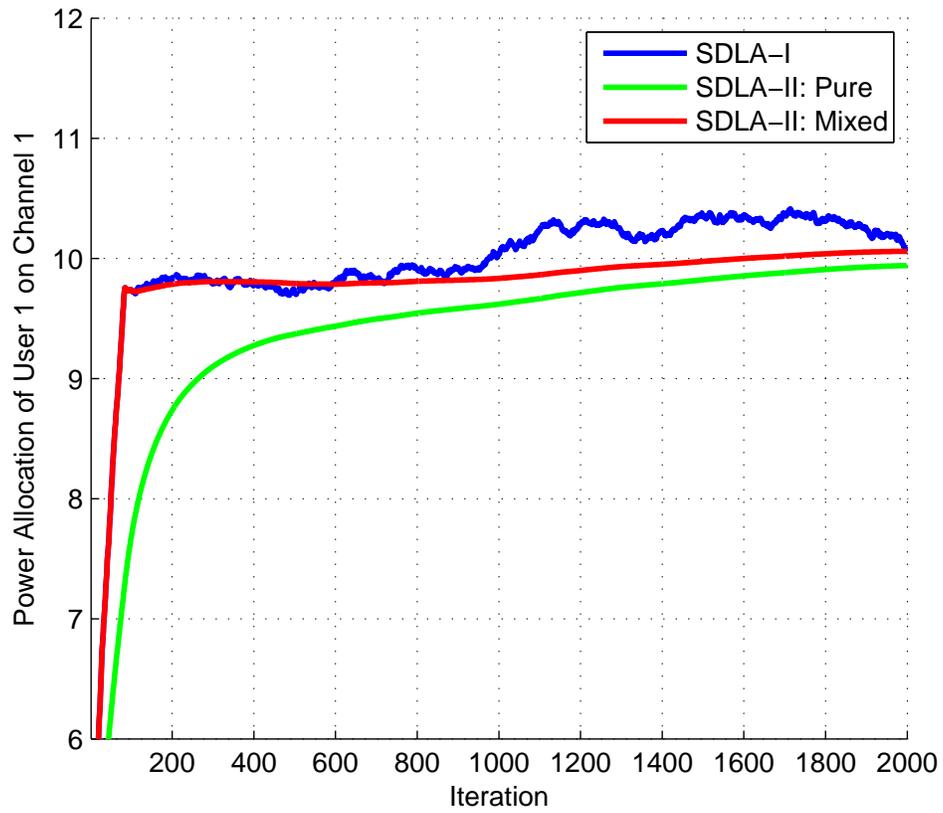}
\caption{Learning NE with Averaging - SDLA-II: $\upsilon = 30\%$ and $a_n = 0.5$.}
\label{fig:5}
\end{figure}

\end{document}